\documentclass[aps,prl,floatfix,superscriptaddress,10pt,twocolumn]{revtex4}
\usepackage{xcolor,amsmath,amssymb,graphicx,latexsym,subfigure}
\usepackage{threeparttable,multirow,txfonts,makecell,tabularx}
\DeclareMathAlphabet{\mathcal}{OMS}{cmsy}{m}{n}
\usepackage{appendix}
\usepackage{hyperref}

\renewcommand{\arraystretch}{1.2}

\newcommand{\eg}{e.g.,\ }	
\newcommand{\ie}{i.e.,\ }	

\newcommand{\TeV}{{\rm TeV}} 
\newcommand{\GeV}{{\rm GeV}} 
\newcommand{\MeV}{{\rm MeV}} 
\newcommand{\DM}{{\rm DM}} 
\newcommand{\DW}{{\rm DW}} 
\newcommand{\GW}{{\rm GW}} 

\newcommand{\beq}{\begin{equation} \begin{aligned}}
\newcommand{\eeq}{\end{aligned} \end{equation}}
\graphicspath{{Figures/}}

\begin{document}
\title{Domain Wall Network: A Dual Solution for Gravitational Waves and Hubble Tension?}

\author{Ligong Bian}
\email{lgbycl@cqu.edu.cn}
\affiliation{Department of Physics and Chongqing Key Laboratory for Strongly Coupled Physics,
Chongqing University, Chongqing 401331, China}
\affiliation{Center for High Energy Physics, Peking University, Beijing 100871, China}

\author{Shuailiang Ge}
\email{sge@pku.edu.cn}
\affiliation{Center for High Energy Physics, Peking University, Beijing 100871, China}
\affiliation{School of Physics and State Key Laboratory of Nuclear Physics and Technology, Peking University, Beijing 100871, China}

\author{Changhong Li}
\affiliation{Department of Astronomy, Key Laboratory of Astroparticle Physics of Yunnan Province, School of Physics and Astronomy, Yunnan University, No.2 Cuihu North Road, Kunming,  650091 China}

\author{Jing Shu}
\email{jshu@pku.edu.cn}
\affiliation{School of Physics and State Key Laboratory of Nuclear Physics and Technology, Peking University, Beijing 100871, China}
\affiliation{Center for High Energy Physics, Peking University, Beijing 100871, China}
\affiliation{Beijing Laser Acceleration Innovation Center, Huairou, Beijing, 101400, China}

\author{Junchao Zong}
\affiliation{Department of Physics, Nanjing University, Nanjing 210093, China}
\affiliation{CAS Key Laboratory of Theoretical Physics, Institute of Theoretical Physics, Chinese Academy of Sciences, Beijing 100190, China}

\begin{abstract}
We explore the possibility that domain wall networks generate the stochastic gravitational wave background (SGWB) observed as a strong common power-law process in the Data Release-2 of Parkes Pulsar Timing Array. We find that a broad range of parameters, specifically wall tension around $\sigma_{\text{DW}} \sim (29-414 , \text{TeV})^3$ and wall-decay temperature within $T_d \sim 20-257 , \text{MeV}$, can explain this phenomenon at a $68\%$ credible level. Meanwhile, the same parameters could ease the Hubble tension if particles from these domain wall networks decay into dark radiation. We establish a direct analytical relationship, $\Omega_{\text{GW}}(f_p,T_0) h^2 \sim \Omega_{\text{rad}} h^2 ( \Omega_\nu \Delta N_{\text{eff}})^2$, to illustrate this coincidence, underlining its importance in the underlying physics and potential applicability to a wider range of models and data. Conversely, if the common power-law process is not attributed to domain wall networks, our findings impose tight limits on the wall tension and decay temperature.

\end{abstract}
\maketitle

\noindent{\it \bfseries  Introduction.} 
A Domain Wall (DW) is a two-dimensional topological defect, forming as a field connects discrete vacua, possibly manifesting during the Universe's early phase transition through the Kibble-Zurek mechanism~\cite{kibble1976topology,zurek1985cosmological}\footnote{For topological defects formation during the phase
transition through Kibble-Zurek mechanism and its connection with dark matter, see Ref.\cite{Murayama:2009nj}.}. The dynamics of a DW network have
nonzero quadrupole momentum and thus can generate gravitational waves (GWs)~\cite{kawasaki2011study, Hiramatsu2013, Hiramatsu2014}, but their slower dilution compared to matter and radiation can dominate the universe’s energy density, posing the Domain Wall problem~\cite{zel1974cosmological}. The application of a biased potential can rapidly decay these walls, thereby marking the cessation of GW production~\cite{sikivie1982axions,gelmini1989cosmology,larsson1997evading}. The domain wall tension and the biased potential influence the magnitude and peak-frequency position of the resultant GW spectrum.

The Stochastic GW Background (SGWB) generated by DW networks falls within the detection range of Pulsar Timing Array (PTA) experiments for certain parameters. PTAs, designed to monitor millisecond pulsars with precise Time of Arrivals, have discovered a common power-law (CPL) process in the analysis of NANOGrav 12.5-year data~\cite{NANOGrav:2020bcs}, further confirmed by other PTA collaborations such as Parkes Pulsar Timing Array (PPTA)~\cite{Goncharov:2021oub} and European Pulsar Timing Array (EPTA)~\cite{Chen:2021rqp}. This CPL signal, while encouraging as potential first light of GW detection via PTAs, necessitates further observation of the ``Hellings-Downs" (HD) correlation for a definitive GW explanation~\cite{Hellings:1983fr}. The signal's origin remains elusive, with investigations into various sources such as supermassive black hole binaries, first-order phase transitions, and cosmic strings~\cite{Bian:2020urb, Xue:2021gyq, Bian:2022tju, NANOGrav:2021flc, Yonemaru:2020bmr, Chen:2022azo}. The SGWB induced by domain walls shows a broken power-law shape, distinguishing it from the wide-frequency-range plateau shape exhibited by cosmic strings.


Besides decaying into GWs, most DW network energy is released as free particles. In this Letter, one intriguing observation is that the free particles can alleviate the Hubble tension if assuming they can further decay into dark radiation \footnote{
A dark radiation solution will increase the effects of Silk damping and neutrino drag (see e.g., Refs.~\cite{Hou:2011ec,
Green:2019glg,
Wallisch:2018rzj,
Follin:2015hya, 
Baumann:2015rya,
Baumann:2017lmt,
Baumann:2017gkg,
Baumann:2019keh}), and the CMB measurements and large-scale structures thus strongly constrain the dark radiation species. The CMB constraints are based on the fact that the Silk damping and the first CMB acoustic peak respond differently to dark radiation which breaks the degeneracy between the Hubble tension $H_0$ and the effective number of relativistic species $N_{\rm eff}$~\cite{Knox:2019rjx}.
We should also point out that the above effects, Silk damping and neutrino drag, can be counteracted to some extent by assuming that the dark radiation is self-interacting~\cite{Schoneberg2021TheModels}.
}. 
Here the Hubble tension refers to the $4.1\sigma$ conflict between the values of the current Hubble constant derived from the early Universe measurements~\cite{Aghanim2020PlanckResults} and from the local late Universe measurements~\cite{Riess:2019cxk}. Remarkably, the same parameter space accounting for the CPL process can simultaneously alleviate the Hubble tension from a stark $4.1\sigma$ to a more reconcilable $2.7\sigma$ (\textit{Planck 2018} $+$ \textit{CMB lensing}, and $3.2\sigma$ with  \textit{BAO} and \textit{Pantheon} data further included~\footnote{We acknowledge and thank the anonymous referee for the reminder to include \textit{BAO} and \textit{Pantheon} data here.}). 
Motivated by this intriguing coincidence in the parameter space, we have developed an analytical relation between $\Delta N_\text{eff}$ and $\Omega_\text{GW}$, Eq.~\eqref{eq:omegadn}. Our analysis reveals that this coincidence naturally emerges from the relation we propose, elucidating the observed fixed decay ratios into gravitational waves and dark radiation within the DW parameter space. 
This coincidence holds true in general attempts of using a DW network to explain the potential SGWB, and it applies to the observation datasets from various PTA collaborations~\cite{NANOGrav:2020bcs, Goncharov:2021oub, Chen:2021rqp, NANOGrav:2023gor,NANOGrav:2023hvm,NANOGrav:2023hde, Antoniadis:2023ott,Xu:2023wog}.




Specifically, in this Letter, we utilize the second dataset released by Parkes PTA (PPTA) to probe a potential SGWB signal \cite{Kerr_2020}. The DW tension and the decay temperature
are considered as two free parameters in our search for corresponding GW signals. We ascertain that specific parameter space of the DW network can account for the CPL process observed in the PPTA data, and the observed Hubble tension can be substantially alleviated by the concurrent production of dark radiation. 
Alternatively, if we consider the CPL process originates from other sources other than a DW network, we can establish stringent constraints on DW parameters.

\noindent{\it \bfseries SGWB spectra from DW networks.}
The interaction between DWs is so efficient that their density after its formation will soon saturate the requirement of causality, \ie approximately one piece of DW per Hubble patch $H^{-3}$. The energy density of DW networks thus evolves with the scaling behavior $\rho_{\DW}(t) \propto H(t)$, which has been found and confirmed in multiple numerical simulations~\cite{Hiramatsu2013,Hiramatsu2014,press1989dynamical,ryden1990evolution,garagounis2003scaling,leite2013accurate}. The DW density is $\rho_{\DW} = \mathcal{A}\sigma_{\DW}H$ where $\sigma_{\DW}$ is the wall tension and $\mathcal{A}$ is the area parameter. DW networks radiate GWs with the power 
 $P_{\GW}\sim G\dddot{Q}_{ij}\dddot{Q}_{ij}$ where $Q_{ij}\sim \mathcal{A} \sigma_{\DW} H^{-4}$ is the quadrupole momentum of the walls. Then the energy density of GWs can be expressed as 
\beq
\label{eq:rhoGW}
\rho_{\GW}= \epsilon P_{\GW}t/H^{-3} =\epsilon G \mathcal{A}^2 \sigma_{\DW}^2 = \frac{3 \epsilon}{8 \pi} \frac{\rho^2_{\DW}}{\rho_{c}}
\eeq
where an efficiency parameter $\epsilon$ is introduced. $\rho_{c} = 3H^2/(8\pi G)$ is the critical density. $\mathcal{A}$ and $\epsilon$ are constant with time in the scaling regime and can be determined by numerical simulations~\cite{Hiramatsu2013,Hiramatsu2014}.

To avoid the DW domination problem, we introduce a biased potential $\Delta V$ to quickly kill the networks~\cite{sikivie1982axions,gelmini1989cosmology,larsson1997evading} by breaking the degeneracy of vacua on the two sides of a wall. Such wall decay happens at 
$H^{-1}(T_d)\simeq \sigma_{\DW}/\Delta V$
when the biased potential energy of a vacuum patch is comparable with the energy of boundary wall. The value of the bias term needs to be adjusted so that the wall decay happens at a specific $T_d$. For an alternative mechanism to destroy DWs and thus solve the DW problem, see \eg Ref.~\cite{Stojkovic_2005}. 
$T_d$ marks the stage that the GW radiation stops, which determines the peak frequency of the GW spectra, $f_p(T_d)\simeq H(T_d)$~\cite{Hiramatsu2013,Hiramatsu2014}. We assume that $T_d$ is in the radiation-dominant era. Then, the peak frequency today is 
\beq\label{eq:fpeak}
f_{p}(T_0) 
&\simeq 1.13\times 10^{-8}{\rm~Hz}\cdot \left[\frac{g_{*}(T_d)}{10.75}\right]^{1/6}\left(\frac{T_d}{100~{\rm MeV}}\right).
\eeq
where $g_{*}$ ($g_{*s}$) is the effective degrees of freedom for energy (entropy). We then define the GW spectra as 
\beq
\label{eq:GWspec}
\Omega_{\GW}(f,T_d) \equiv \frac{1}{\rho_{c}(T_d)}\frac{d\rho_{\GW}(f,T_d)}{d\ln f}
=\frac{d\epsilon}{d\ln f} \frac{1}{\epsilon} \frac{\rho_{\GW}(T_d)}{\rho_{c}(T_d)}. 
\eeq
The present GW spectra at peak frequency is thus 
\beq\label{eq:OmegaPeak}
&\Omega_{\GW}(f_p,T_0) h^2 
=\tilde{g} \cdot \Omega_{\rm rad}h^2 \cdot 
\Omega_{\GW}(f_p,T_d)
\\
& \simeq 6.5\times 10^{-10} \mathcal{A}^2 \tilde{\epsilon} \cdot \left[\frac{10.75}{g_{*}(T_d)}\right]^{4/3}\left(\frac{\sigma_{\DW}}{10^{6}~\TeV^3}\right)^{2}
\left(\frac{100{\rm~MeV}}{T_d}\right)^4
\eeq
where $\tilde{\epsilon}\equiv [d\epsilon/d\ln f]_{f=f_p}$. For convenience, we defined $\tilde{g} = g_{*}(T_d)/g_{*}(T_0) \cdot [g_{*s}(T_0)/g_{*s}(T_d)]^{4/3} \approx 0.83$ which collects the effects of $g$s' changing during cosmic evolution.
The present radiation density parameter is $\Omega_{\rm rad}h^2 = \pi^2/30 \cdot g_{*}(T_0) T_0^4 / [\rho_{c}(T_0)h^2] \simeq 4.15\times 10^{-5}$.

As known, the spectra of $f<f_p$ goes as $\sim f^3$ because of causality~\cite{Caprini_2020,caprini2009general}. For $f>f_p$, simulations show that the slope is close to $\sim f^{-1}$ but only slightly depends on the ``domain wall number" $N_{\DW}$ which is the number of vacua
\cite{Hiramatsu2013,Hiramatsu2014} We consider scenarios of $N_{\DW}\geq 2$ since the $N_{\DW}=1$ string-wall network is unstable and would disappear immediately after their formation without significant GWs production~\cite{Barr:1986hs}. In the main text, we take $\Omega_{\GW}(f,T_0)h^2 = \Omega_{\GW}(f_p,T_0)h^2 \times (f/f_p)^{-1.077}$
 for $f>f_p$ and present the result of the scenario $N_{\DW}=2$. For other scenarios with different $N_{\DW}$, see  appendix for details. 

\begin{figure}[!htp]
    \centering
    \includegraphics[width=\linewidth]{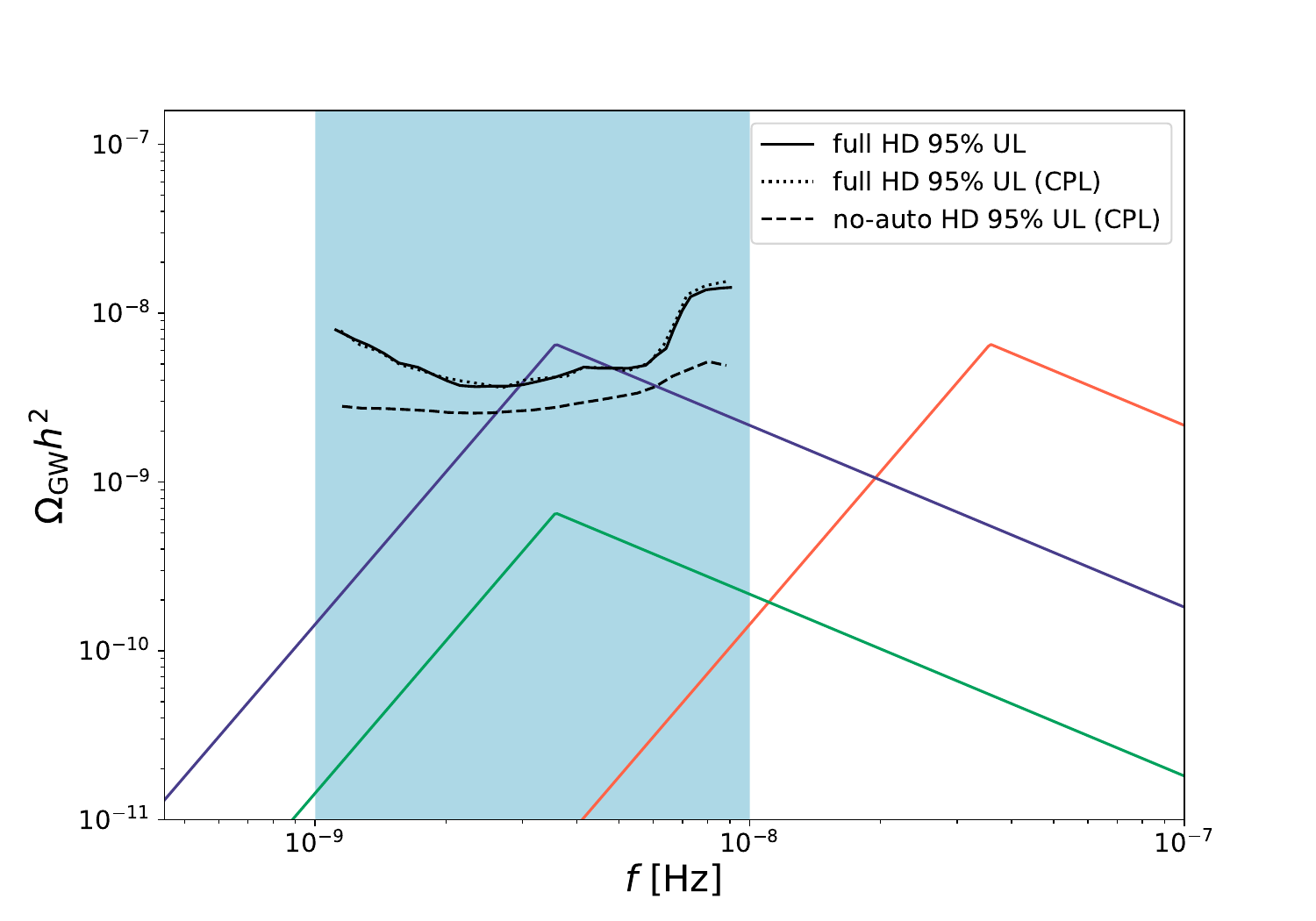}
    \caption{The 95\% upper limit (U.L.) on the SGWB amplitude by searching the 2nd released PPTA data. The shaded region indicates the PPTA-sensitive frequency range.
    The black solid, dotted, and dashed lines are the search results under different hypotheses H2/3/4 in Table~\ref{BYS2} (but with the DW spectra replaced by the free spectra in the corresponding hypotheses). We also show the GW spectra of $N_{\rm DW}=2$ as an example with different values of $( \log_{10}\sigma_{\DW}/\GeV^3,\log_{10}T_{d}/ \GeV)$: $(16.5,-0.5)$ in red, $(14.5,-1.5)$ in purple and $(14.0,-1.5)$ in green for illustrative purpose.
    }
    \label{FS2}
\end{figure}

Throughout this work, we take the parameters $\mathcal{A}=1.2$, $\tilde{\epsilon}=0.7$~\cite{Hiramatsu2013,Hiramatsu2014} and fix $g_{*}(T_d)$ at 10.75. In Fig.~\ref{FS2}, we show the GW spectra for various values of $\sigma_{\DW}$ and $T_{d}$, in comparison with the free-spectra searching results for the SGWB amplitude using the PPTA dataset (2nd release). This plot tells the rough region of DW parameters that the PPTA data is sensitive to. Next, we carry out detailed data analysis employing the second data release of the PPTA data. The details of noise modeling and Bayesian data analysis can be found in the  appendix.



\begin{table*}[!htbp]
    \caption{Description of the hypotheses, Bayes factors, and parameters estimation of $N_{\DW}=2$. ``—" in the column of Parameters Estimation represents for the parameter is absent in that hypothesis.}
    \centering
    \setlength{\tabcolsep}{1.8mm}
    \renewcommand\arraystretch{1.8}
    \footnotesize
\begin{tabular}{|c|c|c|c|c|c|c|c|c|}
\hline
\multicolumn{2}{|c|}{\multirow{2}{*}{Hypothesis}} &\multirow{2}{*}{Pulsar Noise} & \multirow{2}{*}{CPL} & \multirow{2}{*}{SMBHB} & \multirow{2}{*}{DW Spectra} & \multicolumn{3}{c|}{Parameter Estimation ($68\%$ C.L.)}  \\ \cline{7-9}


\multicolumn{2}{|c|}{}  &  &  &  &    & \multicolumn{1}{c|}{$\log_{10}T_d /\GeV$, $\log_{10}\sigma_{\DW}/\GeV^3$} & 
$A_{\text{CPL}}, \gamma_{\text{CPL}},A_{\text{SMBHB}}$ & Bayes Factors 
\\ \hline
\multicolumn{2}{|c|}{H0} & \checkmark &  &  &   &\multicolumn{1}{c|}{—} &\multicolumn{1}{c|}{—} & —\\ \hline
\multicolumn{2}{|c|}{H1}  & \checkmark & \checkmark &  & &  \multicolumn{1}{c|}{—} & $-14.48_{-0.64}^{+0.62}$, $3.34_{-1.53}^{+1.37}$, — &$10^{3.2}$ (/H0)
\\ \hline

\multicolumn{2}{|c|}{\multirow{1}{*}{H2}} & \multicolumn{1}{c|}{\multirow{1}{*}{\checkmark}} &  &  & \multicolumn{1}{c|}{\multirow{1}{*}{\checkmark (full HD)}} &   \multicolumn{1}{c|}{$-1.24_{-0.47}^{+0.65}$, $14.60_{-1.20}^{+2.25}$} & \multicolumn{1}{c|}{\multirow{1}{*}{}} &$10^{2.1}$ (/H0)\\

\hline 

\multicolumn{2}{|c|}{\multirow{1}{*}{H3}} & \multicolumn{1}{c|}{\multirow{1}{*}{\checkmark}} & \multicolumn{1}{c|}{\multirow{1}{*}{\checkmark}} &  & \multicolumn{1}{c|}{\multirow{1}{*}{\checkmark (full HD)}} &  \multicolumn{1}{c|}{$>-2.48$, $<17.73\ $ ($95\%$ C.L.)} & $-14.67_{-0.98}^{+0.65}$, $3.42_{-1.65}^{+1.47}$, — &$1.02$ (/H1)
\\ 

\hline

\multicolumn{2}{|c|}{\multirow{1}{*}{H4}} & \multicolumn{1}{c|}{\multirow{1}{*}{\checkmark}} & \multicolumn{1}{c|}{\multirow{1}{*}{\checkmark}} &  & \multicolumn{1}{c|}{\multirow{1}{*}{\checkmark (no-auto HD)}} &  \multicolumn{1}{c|}{$>-2.57$, $<17.75\ $  ($95\%$ C.L.)} &$-13.62_{-0.14}^{+0.14}$, $4.67_{-0.32}^{+0.33}$, — &$0.89$ (/H1)\\

\hline
\multicolumn{2}{|c|}{H5} & \checkmark &  & \checkmark &  &  \multicolumn{1}{c|}{—} & —, —, $-14.89_{-0.12}^{+0.10}$ &$10^{3.3}$ (/H0)
\\ \hline

\multicolumn{2}{|c|}{\multirow{1}{*}{H6}} & \multicolumn{1}{c|}{\multirow{1}{*}{\checkmark}} & & \multicolumn{1}{c|}{\multirow{1}{*}{\checkmark}} &   \multicolumn{1}{c|}{\multirow{1}{*}{\checkmark (full HD)}} &  \multicolumn{1}{c|}{$>-1.04$, $<16.45$} &  —, —, $-14.92_{-0.16}^{+0.11}$ &$0.90$ (/H5)\\

\hline
\end{tabular}\label{BYS2}
\end{table*}

\noindent{\it \bfseries  Results.}
To interpret PPTA data, we consider several typical hypotheses (labeled by Hn) which are summarized in Table~\ref{BYS2}. H0 only includes the pulsar noise. H1 is to check the existence of a strong CPL signal. H2 is to search for a DW-induced SGWB signal with the full HD correlation included. Since the origin of the CPL signal is still not clear, we further consider hypotheses H3 and H4 where we add the CPL signal as an unknown background in searching for a DW-induced SGWB signal. The difference between H3 and H4 is that H3 includes the full HD correlation while H4 only includes the off-diagonal (\ie no-auto) HD correlation. In addition, we assume the CPL signal has an astrophysical origin, generated by SMBHBs. Analogous to H1 and H3, we get two new hypotheses, H5 and H6. In all cases, the Bayes ephemeris has been taken to include uncertainties from the solar system. 

It turns out that the searches in PPTA data yield similar results for $N_{\DW}=2$ to $6$ DW networks, which is as expected since they have similar spectra as can be seen in Table \ref{fitting} in the  appendix. To be concise, in the following we present the result of $N_{\DW}=2$ case as a benchmark while leaving the details of $N_{\DW}>2$ results in the  appendix.  

\begin{figure}[htp]
    \centering
    \includegraphics[width=1\linewidth]{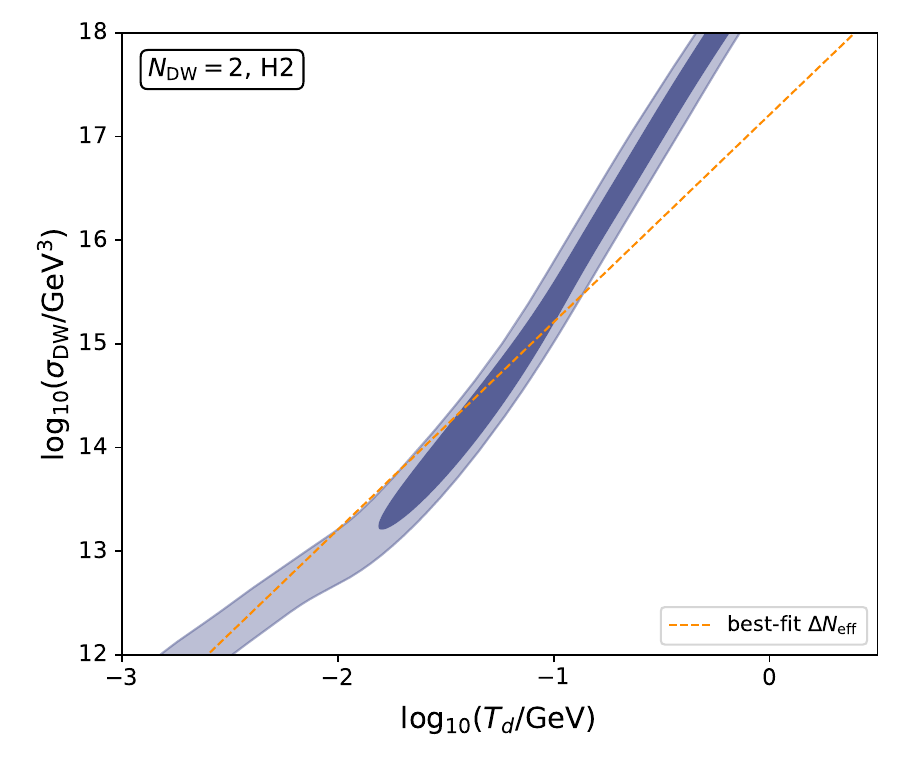}
    \caption{Posterior distribution of $\sigma_{\DW}$ and $T_d$ for $N_{\DW}=2$ with the hypothesis H2 in Table~\ref{BYS2}. 1 and 2-$\sigma$ regions of distribution are shown in light and dark colors. The orange dash line indicates the parameter region corresponding to the best-fit value of $\Delta N_{\rm eff}=0.151$.}
    \label{ndw2H2}
\end{figure}

First of all, a large Bayes factor ${\rm BF}_{10}=10^{3.2}$ of hypothesis H1 against H0 is found in our analyses, implying the existence of a strong CPL signal, which is consistent with PTA collaborations' work~\cite{NANOGrav:2020bcs, Goncharov:2021oub, Chen:2021rqp}. We then test hypothesis H2 where the CPL process is replaced by the DW signal. We derive a comparable Bayes factor of ${\rm BF}_{20}=10^{2.1}$ when the domain wall tension, $\sigma_{\DW}$, and the decay time, $T_d$, are treated as priors in the H2. The fitting results of these parameters, at the 68\% Credible Level (C.L.), are provided in Table~\ref{BYS2}:
\beq\label{eq:prefer-2}
\sigma_{\DW}\sim (29-414~\TeV)^3
,~~~
T_d\sim 20-257~\MeV.
\eeq
The corresponding 1- and 2-$\sigma$ regions of two-dimensional posterior distributions are depicted in Fig. \ref{ndw2H2}. These strip regions should be interpreted as the range of domain wall parameters favored by PPTA data.

If PPTA data indeed indicates the existence of a DW network based on the H2 result, we consider the physical implications in the following discussion. In fact, GWs only carry a minor portion of DW energy, while the majority of the energy is carried by free particles decayed from the walls~\cite{Hiramatsu2013, Hiramatsu2014}. These particles behave as matter with momentum comparable to their mass, as demonstrated by simulations~\cite{Hiramatsu:2012gg, Kawasaki:2014sqa}. To prevent the matter component (composed of these free particles) from dominating the Universe too early and conflicting with BBN, a natural conclusion is that these free particles will further decay into dark radiation via their couplings with a dark sector (see, \eg Refs.~\cite{gonzalez2020ultralight,davoudiasl2020gravitational,berghaus2020thermal, Agrawal:2018vin}). 
More details are shown in the  appendix. This process increases the effective number of relativistic species by $\Delta N_{\rm eff}\equiv \rho_{\rm DR}/\rho_{\nu}^{\rm SM}$ where $\rho_{\rm DR}$ is the energy density of dark radiation, rescaled by $\rho_{\nu}^{\rm SM}$, the energy density of (a single flavor) standard model neutrinos. By assuming the decay of free particles to dark radiation is instant after $T_d$ (\ie $\rho_{\rm DR}(T_d)\simeq \rho_{\DW}(T_d)$), we obtain 
\beq\label{eq:DeltaNeff}
\Delta N_{\rm eff}
& \simeq
\tilde{g} \cdot \Omega_{\nu}^{-1}\cdot \frac{\rho_{\DW}(T_d)}{\rho_{c}(T_d)} 
\\
&\simeq 
0.077  \mathcal{A} \left[\frac{10.75}{g_{*}(T_d)}\right]^{5/6}
\left( \frac{\sigma_{\rm DW}}{10^{6}{\rm ~TeV}^3} \right)
\left(\frac{100{\rm~ MeV}}{T_d}\right)^2.  
\eeq
$\Omega_{\nu}\equiv \rho_{\nu}^{\rm SM}/\rho_{c} \simeq 0.135$ after neutrino decoupling and $e^+ e^-$ annihilation.
A more comprehensive expression for $\Delta N_{\rm eff}$ is presented in the  appendix, where we also construct a succinct model that simultaneously accommodates the bias leading to wall decay and the portal facilitating the further decay of free particles into dark radiation.



It is known that $\Delta N_{\rm eff}$ can help resolve the Hubble tension, a $4.1\sigma$ tension between local and early-time measurements of the Hubble parameter $H_0$~\cite{Bernal:2016gxb, Aghanim2020PlanckResults,Riess:2019cxk,riess2021cosmic, Schoneberg2021TheModels,DiValentino2021In,Vagnozzi2020NewView}. It has been shown that the existence of self-interacting dark radiation, which might be achieved in our case via the coupling with the free particles 
or other new dark couplings, can reduce the Hubble tension to $2.7\sigma$ with the best-fit value $\Delta N_{\rm eff}=0.151$~\cite{Blinov2020InteractingTension, DiValentino2021In}
\footnote{
Different choices of datasets can reduce the Hubble tension to different levels. Here, the $2.7\sigma$ tension is obtained by fitting to the dataset of \textit{Planck 2018} $+$ \textit{CMB lensing}~\cite{Aghanim2020PlanckResults} with a freely-varying self-interacting $\Delta N_{\rm eff}$. The tension would be reduced to $2.9\sigma$~\cite{Blinov2020InteractingTension} with \textit{BAO} data~\cite{alam2017clustering,ross2015clustering,beutler20116df} included and $3.2\sigma$~\cite{Schoneberg2021TheModels} with \textit{Pantheon} data~\cite{scolnic2018complete} further included.
}
.
Then, using the above expression of $\Delta N_{\rm eff}$ obtained in our scenario, we search for the parameter spaces of $\sigma_{\rm DW}$ 
and $T_d$ that give the best-fit values of $\Delta N_{\rm eff}$. The results are shown in Fig.~\ref{ndw2H2}. We see that the best-fit line largely overlaps with the parameter space preferred by PPTA data. Therefore, both the two observations, PPTA and Hubble tension, point to the same range of DW parameters. This might be a clue to the existence of a DW network in the early Universe which decayed around the temperature $T_d\sim 20-257~\MeV$ according to Eq.~\eqref{eq:prefer-2}.

The discovery that a single DW parameter space accounts for both the CPL and Hubble tensions not only illuminates the underlying mechanisms but also extends its relevance to explaining the nano-Hertz SGWB across various PTA studies~\cite{NANOGrav:2020bcs, Goncharov:2021oub, Chen:2021rqp, NANOGrav:2023gor,NANOGrav:2023hvm,NANOGrav:2023hde, Antoniadis:2023ott,Xu:2023wog}. 
As shown in Fig.~\ref{FS2} (or other similar plots~\cite{NANOGrav:2020bcs, Goncharov:2021oub, Chen:2021rqp, NANOGrav:2023gor,NANOGrav:2023hvm,NANOGrav:2023hde, Antoniadis:2023ott,Xu:2023wog} showing the potential SGWB signal), the amplitude of GW spectrum should be $\Omega_{\rm GW} h^2 \gtrsim 10^{-9}$ in $f\sim 10^{-9}{\rm ~to~} 10^{-8}$~Hz  (at least for the first a few frequency bins which are the most important ones). 
Combining Eqs.~\eqref{eq:rhoGW}, ~\eqref{eq:GWspec}, ~\eqref{eq:OmegaPeak}, and~\eqref{eq:DeltaNeff}, we have
\beq 
\label{eq:coincidence}
\frac{\Omega_{\GW}(f_p,T_0) h^2}{\Omega_{\rm rad} h^2} 
= \frac{3 \tilde{g} \tilde{\epsilon}}{8 \pi}\left(\frac{\rho_{\rm DW}(T_d)}{\rho_{c}(T_d)} \right)^2 = \frac{3 \tilde{\epsilon}(\Delta N_{ \rm eff} \Omega_\nu)^2}{8 \pi \tilde{g}} .
\eeq
Therefore, 
\beq 
\label{eq:omegadn}
\Omega_{\GW}(f_p,T_0) h^2 
&\simeq 2.5\times 10^{-9} \tilde{\epsilon}
\left(\frac{\Delta N_{\rm eff}}{0.151}\right)^2,
\eeq
which suggest the ``coincidence" that if the resultant dark radiation alleviates the Hubble tension, the quadrupole momentum generated by this excited source of dark radiation coincides precisely with the stochastic gravitational wave background (SGWB) as observed today.
The coincidence is two-fold: firstly, it is independent of DW parameters; secondly, the coincidence itself predicts the magnitude of the SGWB we currently observe, which is $\Omega_{\GW}(f_p,T_0) h^2 \sim \Omega_{\rm rad} h^2  ( \Omega_\nu \Delta N_{\rm eff})^2$ with the prefactors omitted. Remarkably, when redshift dilution $\Omega_{\rm rad} h^2$ is factored in to reduce the SGWB $\sim ( \Omega_\nu \Delta N_{\rm eff})^2$ to its current magnitude, this matches the observations reported by various PTA collaborations. Therefore, the two seemingly unrelated topics, the nano-Hertz SGWB and Hubble tension, are closely connected together via this relation. They can be simultaneously addressed by a DW network that once existed. The coincidence can be understood to some extent by knowing the fact that SGWB as radiations themselves would also contribute to $N_{\rm eff}$, as estimated by the nucleosynthesis bound, $\rho_{\rm GW,0}/\rho_{\rm rad,0}\sim 0.227(N_{\rm eff}-3)$ \cite{Maggiore_2000,allen1996stochastic}. It should also be pointed out that it is challenging to fully resolve the Hubble tension by introducing an extra $\Delta N_{\rm eff}$ (see more detailed analysis in Ref.~\cite{Guo:2018ans}). Moreover, a side-effect is that 
the extra $\Delta N_{\rm eff}$ will increase the $\sigma_8$ tension, i.e., the disagreement between CMB measurements and astrophysical measurements on the value of $\sigma_8$ which characterizes the matter perturbation amplitude~\cite{Guo:2018ans}.

\begin{figure}[htp]
    \centering
        \centering
        \includegraphics[width=1\linewidth]{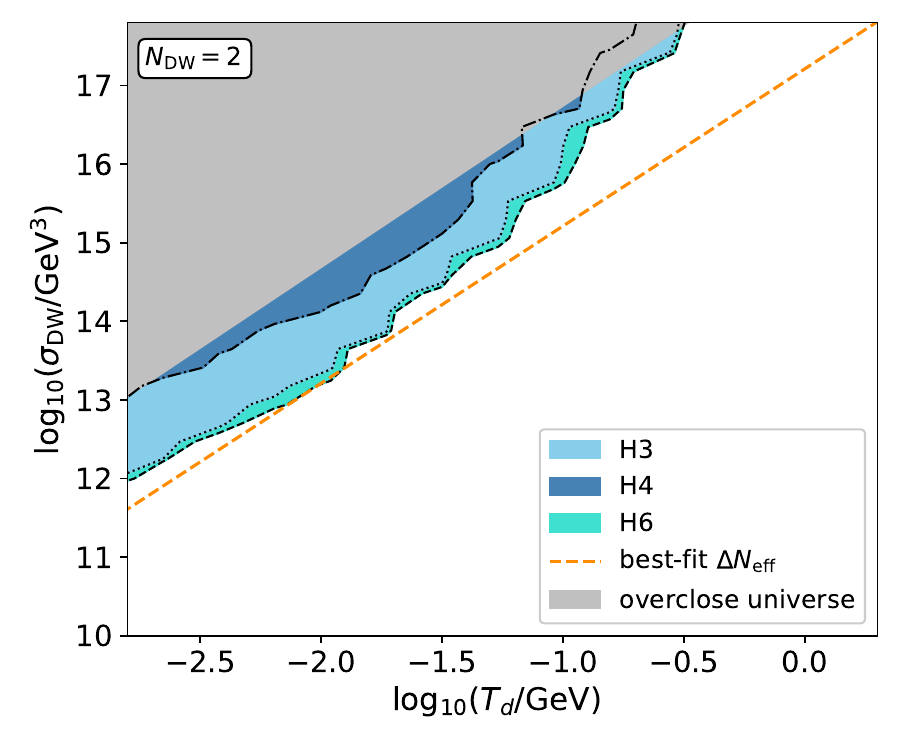}
        \label{fig:ndw2H6}
    \caption{
    95\% C.L. exclusion contour of $\sigma_{\DW}$ and $T_d$  for $N_{\DW}=2$ under hypotheses H3, H4 and H6 in Table~\ref{BYS2}.}
    \label{2hn}
\end{figure}

On the other hand, if the strong CPL signal is not due to DWs, we can set constraints on the DW parameters. 
Alternatively, we specify the unknown CPL signal as generated by SMBHBs. The Bayes factor for H6 against H5 is also less than 1. Both H6 and H3 give similar results as we can see from Table~\ref{BYS2} and Fig.~\ref{2hn}. This means that assuming a specific origin (SMBHB) for the CPL signal essentially does not affect setting exclusion regions for DW parameters. 
As shown in Fig.~\ref{2hn}, roughly the following parameter space in the PPTA-sensitive range is excluded:
\beq\label{eq:exclude-2}
\left(\frac{\sigma_{\DW}}{10^{7} ~ \TeV^3}\right) \gtrsim \left(\frac{T_d}{100 ~\MeV}\right)^2
,~
{\rm for}
~
T_d \sim (1 {\rm -} 10^{2.5}) ~\MeV.
\eeq
It is about one order of magnitude better than the existing constraint from the requirement that DWs should not overclose the Universe~\cite{Saikawa_2017}.

\noindent{\it \bfseries  Summary and discussions.} We find that the same parameter space of DWs that explains the nHz SGWB can also alleviate the Hubble tension from the reported $4.1\sigma$ to a low level, $2.7\sigma$ (\textit{Planck 2018} $+$ \textit{CMB lensing}, and $3.2\sigma$ with  \textit{BAO} and \textit{Pantheon} data further included). This is achieved with the assumption that DW particles can further convert into self-interacting dark radiation.
To shed light on the dual role of DWs,
we have established a direct analytical relation between $\Delta N_\text{eff}$ and $\Omega_\text{GW}$ induced by DWs (see Eq.~\eqref{eq:omegadn}).
This finding holds relevance for the broader range of PTA collaboration studies~\cite{NANOGrav:2020bcs, Goncharov:2021oub, Chen:2021rqp, NANOGrav:2023gor,NANOGrav:2023hvm,NANOGrav:2023hde, Antoniadis:2023ott,Xu:2023wog}. Specifically, we have carried out detailed data analysis based on the PPTA DR2, which has verified the dual role of DWs (see Fig.~\ref{ndw2H2}). 
Therefore, the two seemingly unrelated subjects, nano-Hertz SGWB and Hubble tension, turn out to be closely related within the DW framework. We conclude that a DW network
in the early Universe might simultaneously account for the nHz SGWB and Hubble tension. This might represent a clue to a DW network that ever existed.

For the QCD axion case, we can map the preferred DW parameter space, Eq.~\eqref{eq:prefer-2}, to the axion mass $m_a \sim 10^{-13}-10^{-8}\ {\rm eV}$
\footnote{Lighter scalars might be excluded due to the absence of invisible Higgs decay~\cite{Stojkovic:2013ppa}. Fortunately, the Higgs boson's decay rate into axions under study is nearly negligible, implying minimal impact on the branching fraction of the Higgs into Standard Model particles~\cite{Oikonomou:2023bah}.
}. 
We have used $\sigma_{\rm DW}=9.23 f_a^2 m_a$~\cite{Hiramatsu2013} and the fixed relation $\chi_a = m_a^2 f_a^2 \approx (75.6 \MeV)^4$~\cite{borsanyi2016calculation} where $\chi_a$ is the QCD topological susceptibility and $f_a$ is the decay constant.
Although for such axions to decay into dark radiation (\eg dark photons) may face a difficulty that the axion couplings are usually suppressed by $f_a$,  the coupling with dark radiation could be made large enough via careful model buildings (see \eg Refs.~\cite{Agrawal:2017eqm, Agrawal:2018vin}) to make a fast decay.
Many terrestrial experiments aim to directly search for axions in this mass range, such as ABRACADABRA~\cite{Ouellet_2019,Salemi_2021}, DM-Radio~\cite{DM-Radio}, SRF~\cite{Berlin_2021}, WISPLC~\cite{Zhang_2022}, SHAFT~\cite{Gramolin_2020}, and CASPEr~\cite{JacksonKimball:2017elr}. Searching for GW signal induced by the (axionic) DW network in PPTA data could be a cross-check and complementary to these experiments. We also explored additional hypotheses, setting the CPL process as an unknown background or from SMBHBs.
Under these scenarios, we obtain the parameter space Eq.~\eqref{eq:exclude-2} ruled out at 95\% C.L.
Furthermore, our results indicate that networks with varying $N_{\DW}$ yield similar outcomes.

As this study concludes, we find it thrilling that the latest 15-year data from NANOGrav~\cite{NANOGrav:2023gor,NANOGrav:2023hvm,NANOGrav:2023hde}, EPTA's~\cite{Antoniadis:2023ott} second, and CPTA's~\cite{Xu:2023wog} first data releases all furnish supportive evidence for the HD correlation. However, the HD signature cross-correlation in PPTA's~\cite{Reardon:2023gzh} third data release appears relatively weak. Despite not being able to claim the discovery of nHz SGWB at present, we remain hopeful. Future global PTA data and enhanced large-scale cosmological defect simulations are anticipated to illuminate early universe mysteries, particularly those linked to GWs and DWs.



\noindent{\it \bfseries  Acknowledgements-} This work is supported by the National Key Research and Development Program of China under Grant No. 2020YFC2201501 and 2021YFC2203004. L.B. is supported by the National Natural Science Foundation of China (NSFC) under Grants No. 12075041 and No. 12147102. 
S.G. is supported by NSFC under Grant No. 12247147, the International Postdoctoral Exchange Fellowship Program, and the Boya Postdoctoral Fellowship of Peking University. 
C.L. is supported by the NSFC under Grants No.11963005, and No. 11603018, by Yunnan Provincial Foundation under Grants No.2016FD006 and No.2019FY003005, by Reserved Talents for Young and Middle-aged Academic and Technical Leaders in Yunnan Province Program, by Yunnan Provincial High level Talent Training Support Plan Youth Top Program, and by the NSFC under Grant No.11847301 and by the Fundamental Research Funds for the Central Universities under Grant No. 2019CDJDWL0005. J.S. is supported by Peking Uniersity under startup Grant No. 7101302974 and the National Natural Science Foundation of China under Grants No. 12025507, No.12150015; and is supported by the Key Research Program of Frontier Science of the Chinese Academy of Sciences (CAS) under Grants No. ZDBS-LY-7003. 

\newpage

\appendix
\noindent{\it \bfseries  Appendix}

\section{GW spectra}
Refs.~\cite{Hiramatsu2013, Hiramatsu2014} have numerically simulated in detail the generation of GWs by DW networks with different $N_{\DW}$ based on QCD axion model and $\lambda \phi^4$ model. But their result is quite general as the shape of the DW-induced GWs is primarily model-independent. Here, we directly fit for the slope-power $P$ from the simulation result shown as Figure 6 in \cite{Hiramatsu2013}. The fitted power values are summarized in Table~\ref{fitting}. For the different $N_{dw}$,
the complete GW spectra can be parameterized as
$\Omega_{\GW}(f,T_0)h^2 = \Omega_{\GW}(f_p,T_0)h^2 \times (f/f_p)^n$
where $n=3$(and $n=P$) for $f<f_p$ (and $f>f_p$). 

\begin{table}[!htp]
    \centering
    \caption{Fitted power value $P$ for $f>f_p$.}
    \begin{tabular}{|c|c|c|c|c|c|}
    \hline
     $N_{\DW}$ & 2 & 3 & 4 & 5 & 6 \\
    \hline
     $P$ & -1.077 & -0.972 & -0.887 & -0.807 & -0.731 \\
    \hline
    \end{tabular}\label{fitting}
\end{table}
$n=P$ for $f>f_p$. 

\section{Parameters summary}

The prior distributions and descriptions of the parameters of DW, PTA noise models, and SGWB model are summarized in Table~\ref{PPT}.

\begin{table*}[htp!]
\caption{Parameters and their prior distribution in data analysis. U and log-U stand for the uniform and log-uniform distribution.}
\label{PPT}
\centering
\setlength{\tabcolsep}{2.5mm}
\begin{tabular}{lccc}
\hline\hline
parameter            & \multicolumn{1}{c}{Description}                      & \multicolumn{1}{c}{Prior}                         & \multicolumn{1}{c}{Comments}                 \\ \hline
\multicolumn{4}{c}{Noise parameters($\boldsymbol{\vartheta}$)}                                                                                                                                       \\
EFAC                 & \multicolumn{1}{c}{White-noise modifier per backend} & \multicolumn{1}{c}{U $[0, 10]$}                 & \multicolumn{1}{c}{Fixed for setting limits} \\
EQUAD                & \multicolumn{1}{c}{Quadratic white noise per backend} & \multicolumn{1}{c}{log-U $[-10, -5]$}           & \multicolumn{1}{c}{Fixed for setting limits} \\
ECORR                & \multicolumn{1}{c}{Correlated-ToAs white noise per backend} & \multicolumn{1}{c}{log-U $[-10, -5]$}           & \multicolumn{1}{c}{Fixed for setting limits} \\

$A_{\textrm{SN}}$             & \multicolumn{1}{c}{Spin-noise amplitude}             & \multicolumn{1}{c}{log-U $[-20, -6]$ (search)} & \multicolumn{1}{c}{One parameter per pulsar} \\
                     & \multicolumn{1}{c}{}                                 & \multicolumn{1}{c}{U [$10^{-20}, 10^{-6}$] (limit)}  & \multicolumn{1}{c}{}                         \\
$\gamma_{\textrm{SN}}$        & Spin-noise spectral index                            & U $[0, 10]$                                      & One parameter per pulsar                     \\
$A_{\textrm{DM}}$             & DM-noise amplitude                                   & log-U $[-20, -6]$ (search)                    & One parameter per pulsar                     \\
                     &                                                      & \multicolumn{1}{c}{U [$10^{-20}, 10^{-6}$] (limit)}                                      &                                              \\
$\gamma_{\textrm{DM}}$        & DM-noise spectral index                            & U $[0, 10]$                                      & One parameter per pulsar                     \\

$A_{\textrm{BAND}}$           & Band-noise amplitude 
                  & log-U $[-20, -6]$ (search)                    & One parameter partial pulsars                     \\
                     &                                                      & \multicolumn{1}{c}{U [$10^{-20}, 10^{-6}$] (limit)}                                      &                                              \\
$\gamma_{\textrm{BAND}}$           & Band-noise spectral index
                  &  U $[0, 10]$                    & One parameter  partial pulsars                     \\                
          & One parameter per pulsar                     \\
$A_{\textrm{CHROM}}$             & Chromatic-noise amplitude                                   & log-U $[-20, -6]$ (search)                    & One parameter partial pulsars                     \\
                     &                                                      & \multicolumn{1}{c}{U [$10^{-20}, 10^{-6}$] (limit)}                                      &                                              \\
$\gamma_{\textrm{CHROM}}$        & Chromatic-noise spectral index                            & U $[0, 10]$                        & One parameter partial pulsars                     \\
$n_{\textrm{CHROM}}$        & Index of chromatic effects                            & U $[0, 6]$                        & Fixed for single pulsar                     \\
$A_{\textrm{CPL}}$             & CPL process amplitude                                   & log-U $[-18, -11]$ (search)                    & One parameter per PTA                     \\
                     &                                                      & \multicolumn{1}{c}{U [$10^{-18}, 10^{-11}$] (limit)}                                      &                                              \\
$\gamma_{\textrm{CPL}}$             & CPL process power index                                   & U $[0, 7]$                     & One parameter per PTA                     \\

\hline                    
\multicolumn{4}{c}{DW networks signal parameters ($\boldsymbol{\psi}$)}                                                                                                                                      \\
$N_{\DW}$  & DW number                                       & $[2,3,4,5,6]$ (fixed for each search)                      & One parameter per PTA                        \\


$\sigma_{\DW} [\GeV^3]$  & Wall tension                                       & log-U $[10, 18]$ (search)                      & One parameter per PTA                        \\
                     &                                                      & U [$10^{10}, 10^{18}$] (limit)               &                                              \\

$T_d [\GeV]$  & Decay temperature                                       & log-U $[-3, 5]$ (search)                      & One parameter per PTA                        \\
                     &                                                      & U [$10^{-3}, 10^{5}$] (limit)               &                                              \\
\hline
\multicolumn{4}{c}{BayesEphem parameters ($\boldsymbol{\phi}$)}                                                                                                                               \\
$z_{\rm drift}$          & Drift-rate of Earth’s orbit about ecliptic $z$-axis & U$[-10^{-9}, 10^{-9}]\ {\rm rad}\ {\rm yr}^{-1}$             & One parameter per PTA                        \\
$\Delta M_{\rm Jupiter}$ & Perturbation of Jupiter's mass                       & $N(0,1.5\times 10^{-11})~M_{\bigodot}$          & One parameter per PTA                        \\
$\Delta M_{\rm Saturn}$  & Perturbation of Saturn's mass                        & $N(0, 8.2\times 10^{-12})~M_{\bigodot}$          & One parameter per PTA                        \\
$\Delta M_{\rm Uranus}$  & Perturbation of Uranus' mass                         & $N(0, 5.7\times 10^{-11})~M_{\bigodot}$          & One parameter per PTA                        \\
$\Delta M_{\rm Neptune}$ & Perturbation of Neptune's mass                       & $N(0, 7.9\times 10^{-11})~M_{\bigodot}$          & One parameter per PTA                        \\
$PCA_i$                  & Principal components of Jupiter's orbit              & U $[-0.05, 0.05]$                                          & One parameter per PTA                        \\
\hline\hline
\end{tabular}
\end{table*}

\begin{figure}[!htp]
    \centering
    \includegraphics[width=9.3cm]{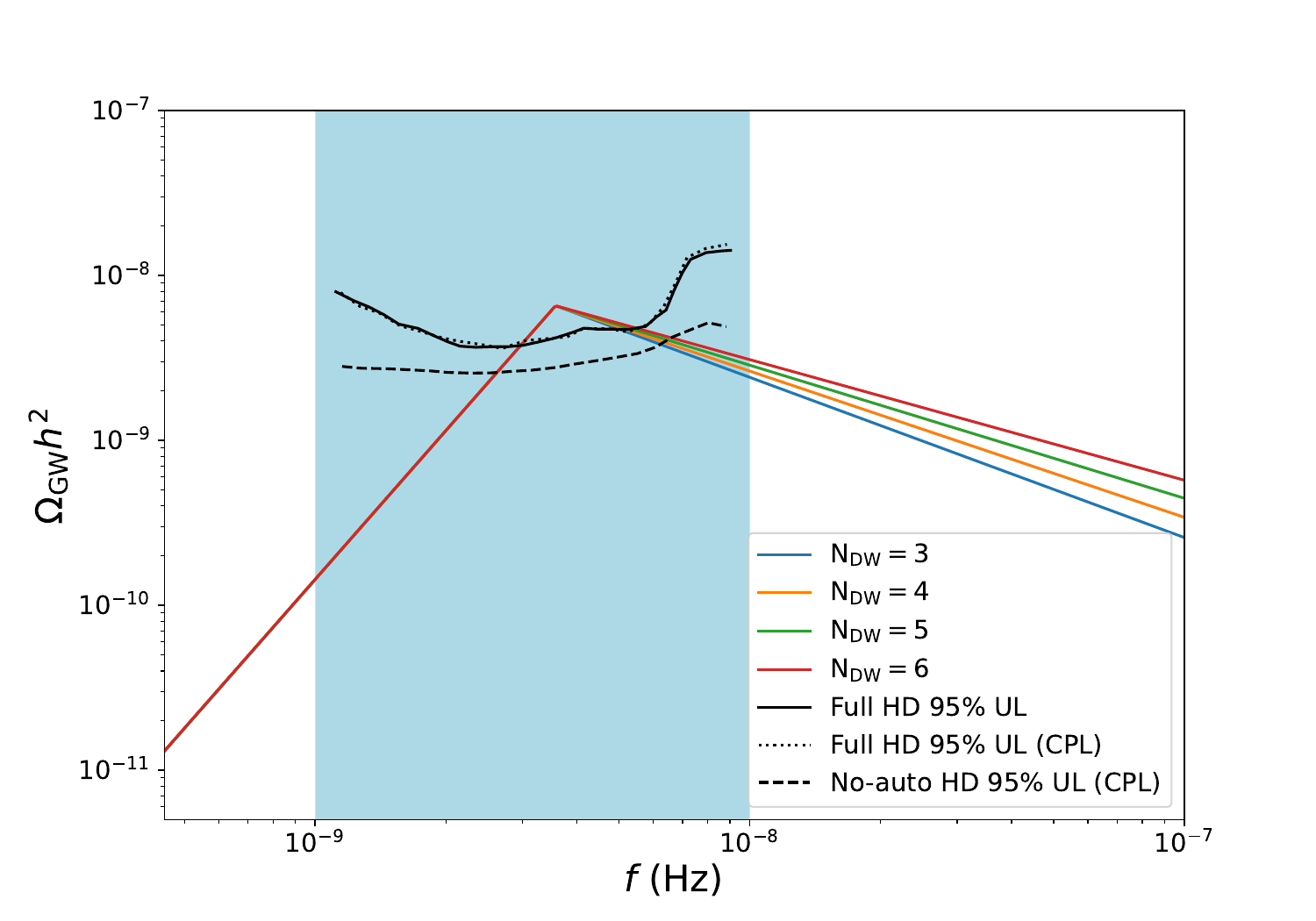}
    \caption{
    Same as Fig.1 in the main text, but for $N_{\DW}\ge2$ cases. Also, the parameters are fixed at $(\log_{10}\sigma_{\DW}/\GeV^3, \log_{10}T_d/ \GeV) = (14.5, -1.5)$ for illustrative purpose.
    }
    \label{FS}
\end{figure}

\section{Data analysis}
Our analysis employs the second data release of the PPTA data.
We adhere to the standard procedure, using the TEMPO2 tool \cite{Hobbs:2006cd,Edwards:2006zg} to fit for the timing model of the pulsar arrays, along with the ENTERPRISE~\footnote{https://github.com/nanograv/enterprise}, ENTERPRISE EXTENSIONS~\footnote{https://github.com/nanograv/enterprise\_extensions}, and PTMCMCSampler~\cite{justin_ellis_2017_1037579} packages for noise modeling and Bayesian analysis. The identification of the SGWB signal in PTA involves searching for the Hellings-Downs (HD) correlation of the time of arrivals (ToAs) for the pulsar array, which necessitates the removal of noise generated during observation. Generally, two types of noises need to be considered: red and white noises. Red noises can originate from irregular pulsar spin (spin noise) or dispersion measure noise of the pulse during transit through interstellar media. 
All types of red noise are modeled to have power-law spectra, parameterized by an amplitude $A$, and a power $\gamma$. We employ EFAC (Error FACtor) and EQUAD (Error added in QUADrature) to account for additional white noise from the correlation of ToAs across multiple frequency channels. This white noise includes ToA uncertainties and ECORR (Error of CORRelation between ToAs in a single epoch) that are not accounted for. In practice, these white noise parameters are fixed by their maximum-likelihood values obtained from single pulsar analysis, as they bear little relation to SGWB model parameters~\cite{Goncharov:2020krd}. All the parameters and their priors used in the Bayesian analysis are summarized in Table \ref{PPT} in the appendix, where we have also considered the constraints from \eg big bang nucleosynthesis (BBN) to set the prior range of DW parameters.


Then, to extract signal information from posterior data, we adopt Bayesian analysis method here. We use the Bayes factor $\rm BF_{10}$ to measure the strength of the signal hypothesis H1 against the null hypothesis H0, which is given by the Savage Dickey formula\cite{10.2307/2958475}
\begin{equation}
{\rm BF}_{10} = \frac{P_{1}(\boldsymbol{D})}{P_{0}(\boldsymbol{D})} = \frac{P(\boldsymbol{\phi}=\boldsymbol{\phi_0})}{P(\boldsymbol{\phi}=\boldsymbol{\phi_0}|\boldsymbol{D})} ,
\end{equation}
where $P_{0/1}(\boldsymbol{D})$ is the evidence of the noise/signal hypothesis with $\boldsymbol{D}$ being the observational data. $\boldsymbol{\phi}$ is the parameters of the signal model, and 
$\boldsymbol{\phi_0}$ is the values making the signal hypothesis null (for instance, the values that make the amplitude of SGWB signal vanish).
By employing this formula, ${\rm BF}_{10}$ can be simply expressed by the ratio of the prior to the posterior probabilities of the null hypothesis. Note that the ``null" and ``signal" are relative concepts depending on the parameters we focus on. 

\section{Results from other hypotheses}

\begin{figure*}[htp!]
    \centering
    \begin{minipage}{0.48\linewidth}
        \centering
        \includegraphics[width=8.5cm]{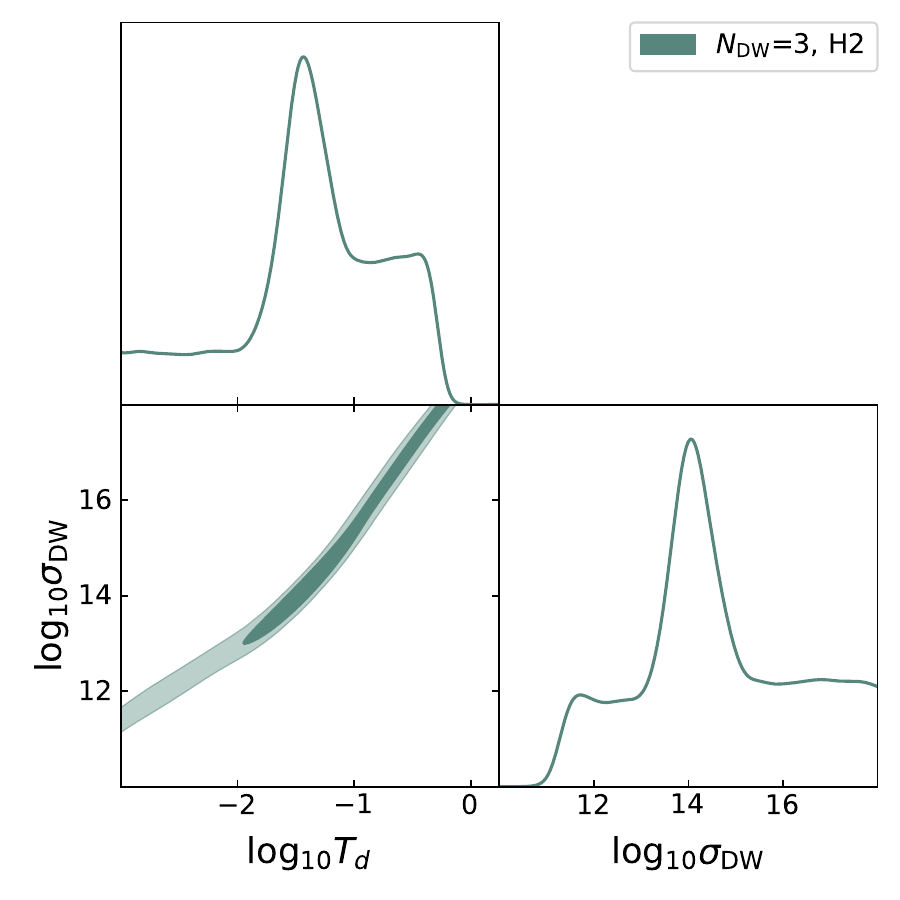}
    \end{minipage}
    \begin{minipage}{0.48\linewidth}
        \centering
        \includegraphics[width=8.5cm]{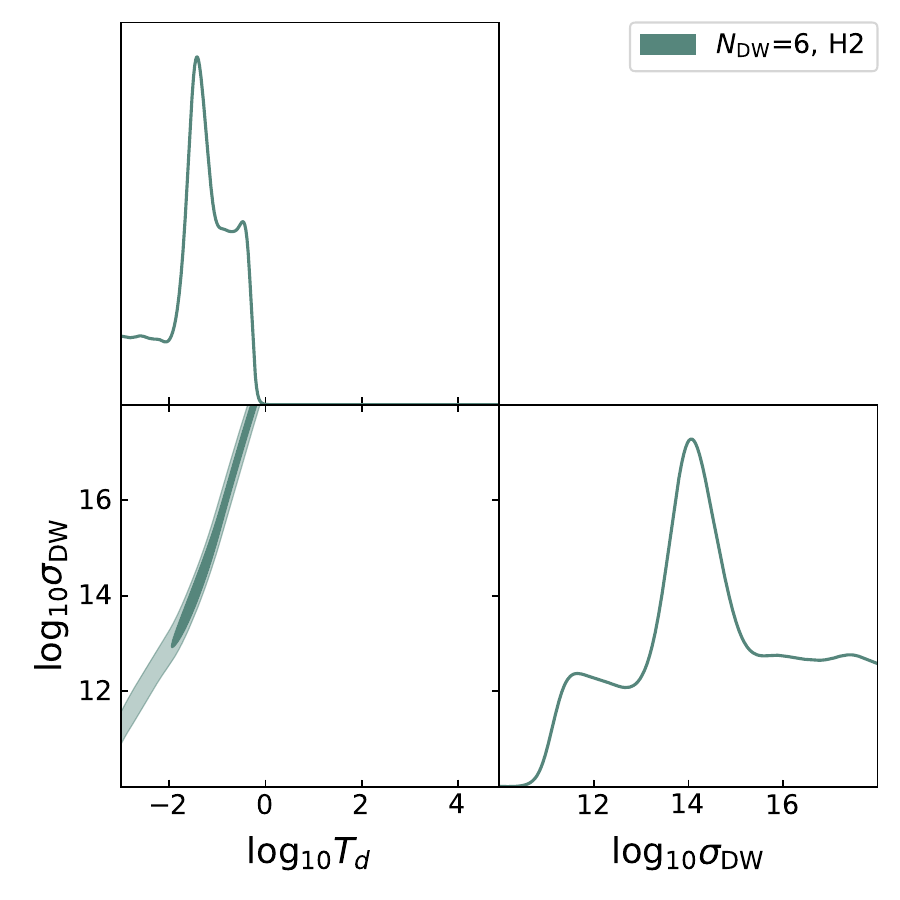}
    \end{minipage}
    \caption{Posterior distribution of the decay temperature $\log_{10}(T_d/\GeV)$ and the wall tension $\log_{10}(\sigma_{\DW}/\GeV^3)$ with  $N_{\DW}=3$ and $6$ for hypothesis H2 in Table \ref{ABYS}. }
    \label{fig:ndwH2_td}
\end{figure*}

\begin{figure*}[ht!]
    \centering
    \begin{minipage}{0.49\linewidth}
        \centering
        \includegraphics[width=8.5cm]{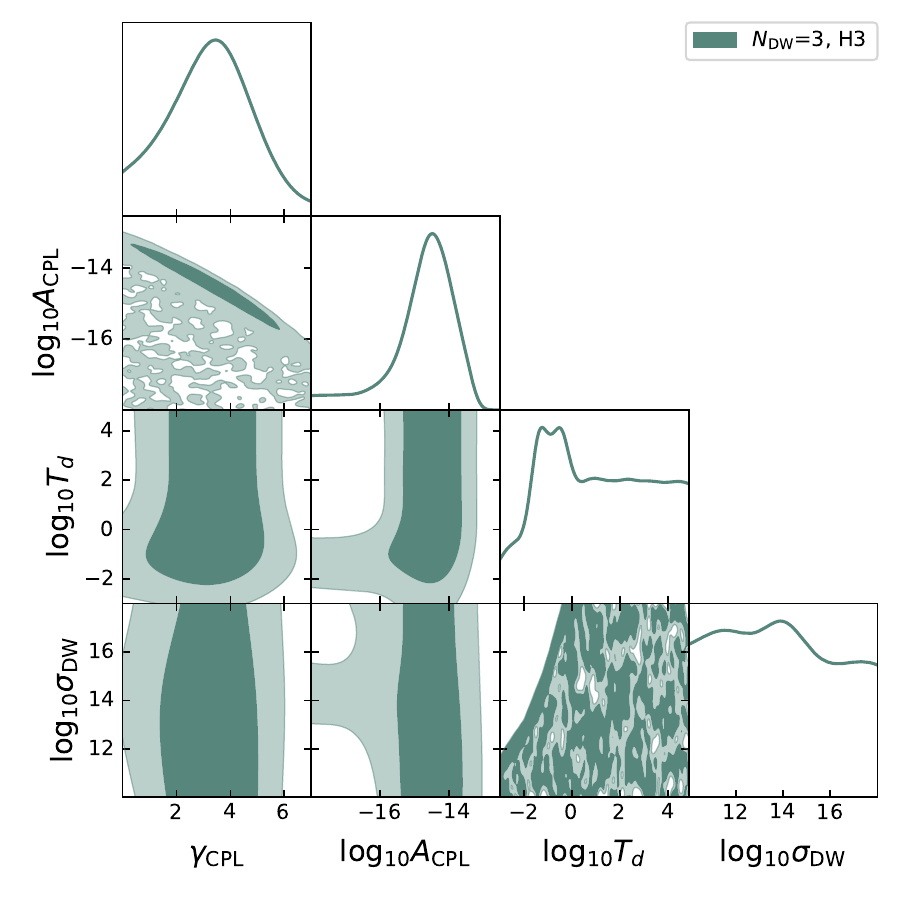}
    \end{minipage}
    \begin{minipage}{0.49\linewidth}
        \centering
        \includegraphics[width=8.5cm]{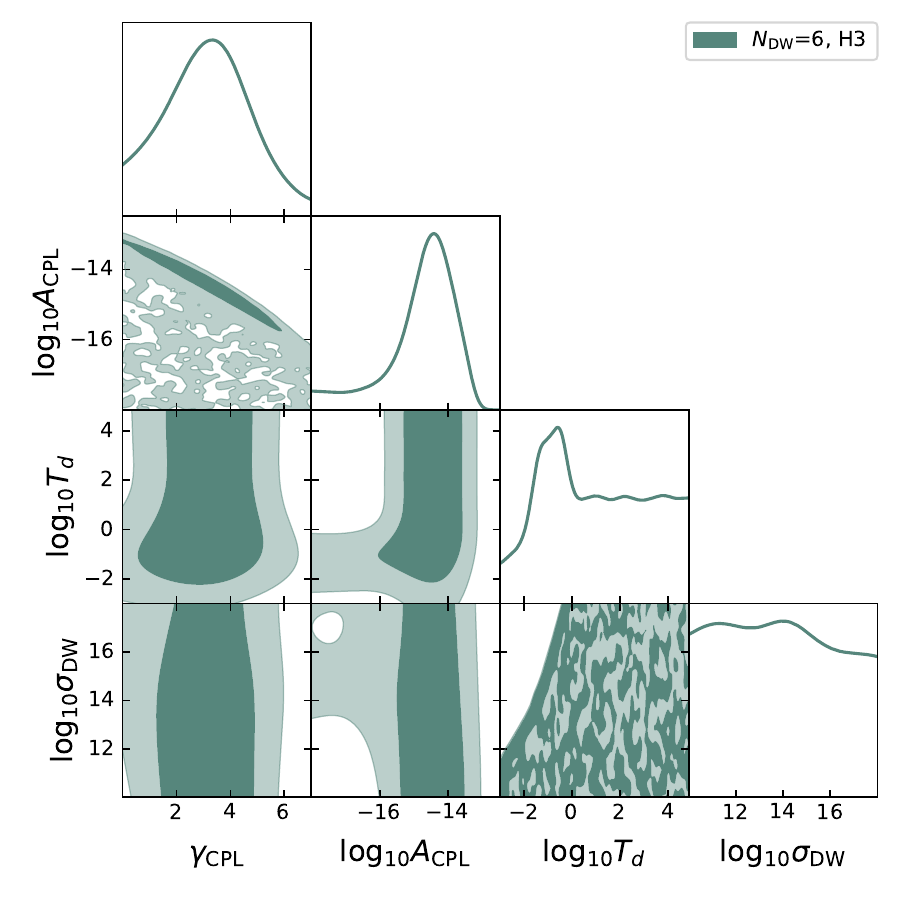}
    \end{minipage}    
    \caption{Posterior distribution of the decay temperature $\log_{10}(T_d/\GeV)$ and the wall tension $\log_{10}(\sigma_{\DW}/\GeV^3)$ with  $N_{\DW}=3$ and $6$ for hypothesis H3 in Table \ref{ABYS}. }
    \label{fig:ndwH3_td}
\end{figure*}

\begin{figure*}[ht!]
    \centering
    \begin{minipage}{0.49\linewidth}
        \centering
        \includegraphics[width=8.5cm]{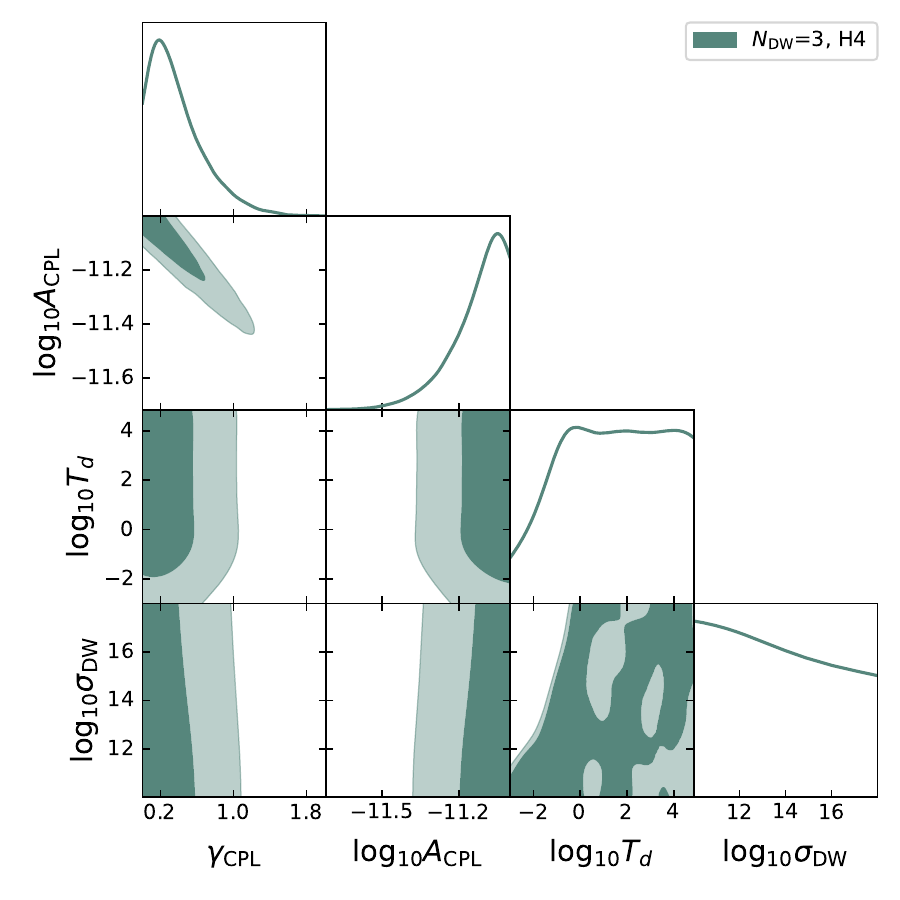}
    \end{minipage}
    \begin{minipage}{0.49\linewidth}
        \centering
        \includegraphics[width=8.5cm]{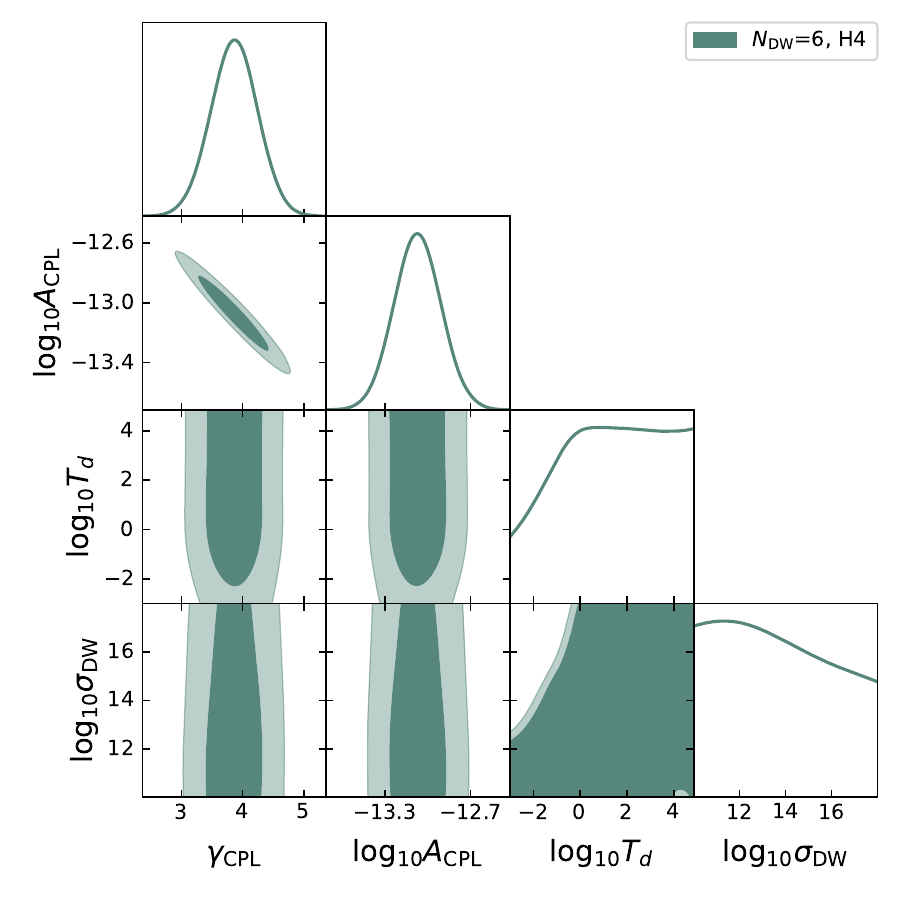}
    \end{minipage}
    \caption{Posterior distribution of the decay temperature $\log_{10}(T_d/\GeV)$ and the wall tension $\log_{10}(\sigma_{\DW}/\GeV^3)$ with $N_{\DW}=3$ and $6$ for hypothesis H4 in Table \ref{ABYS}. }
    \label{fig:ndwH4_td}
\end{figure*}

\begin{figure*}[ht!]
    \centering
    \begin{minipage}{0.49\linewidth}
        \centering
        \includegraphics[width=8.5cm]{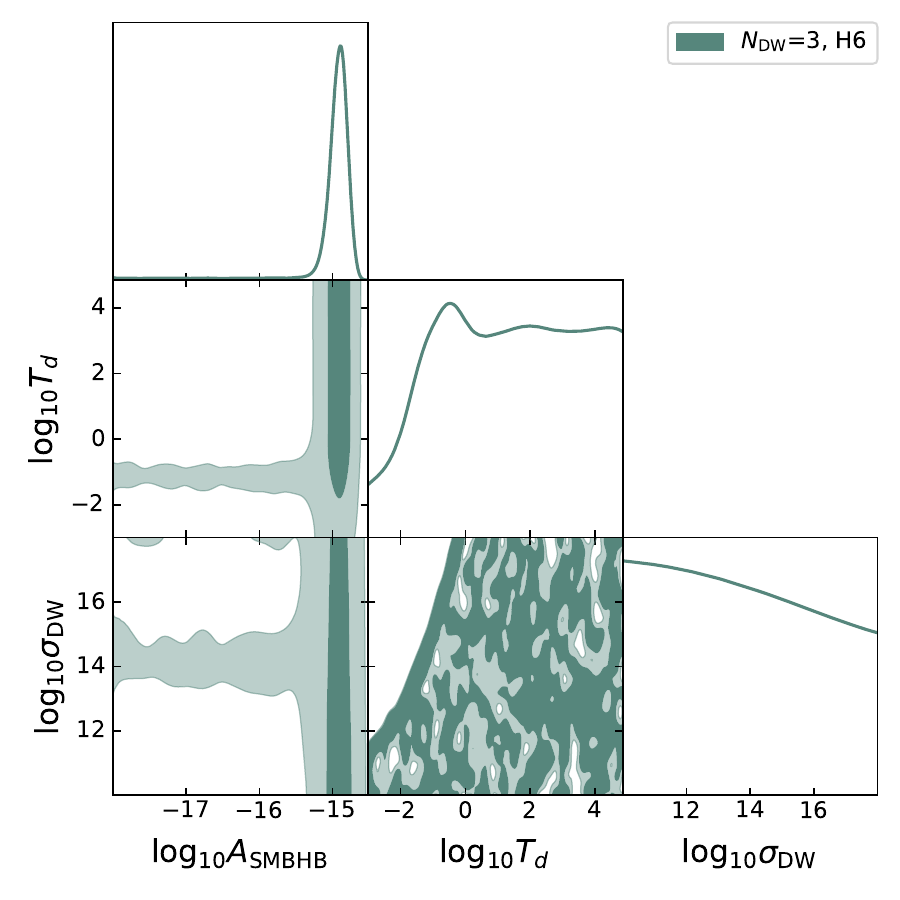}
    \end{minipage}
    \begin{minipage}{0.49\linewidth}
        \centering
        \includegraphics[width=8.5cm]{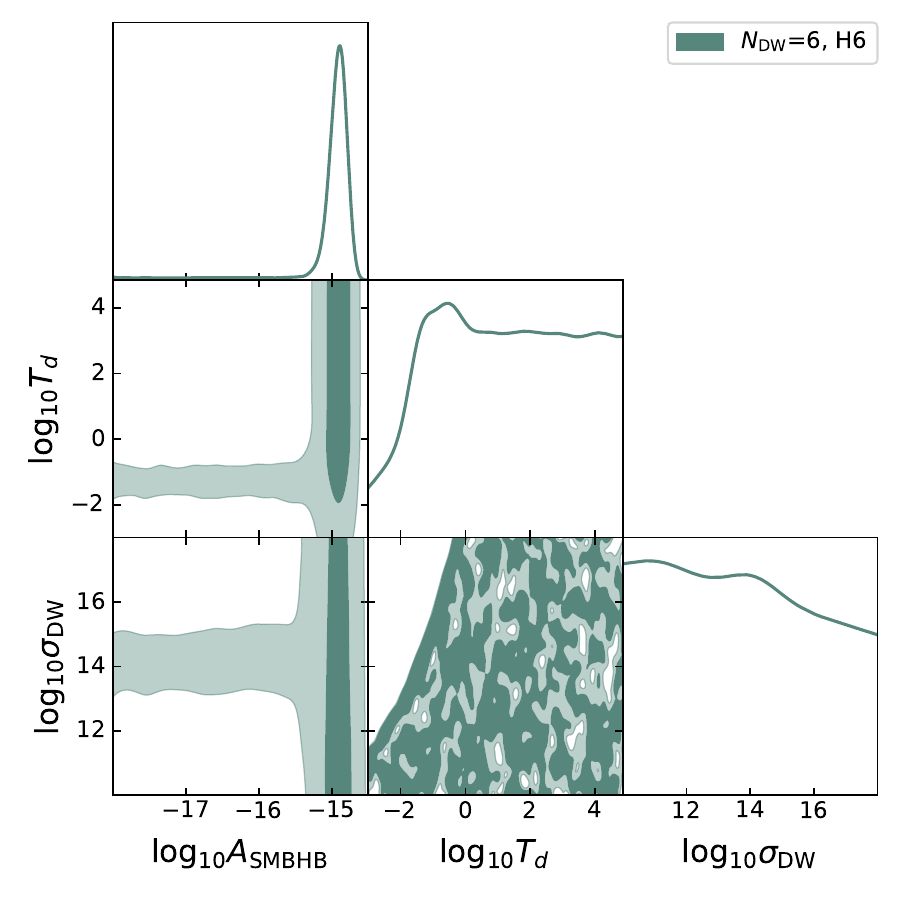}
    \end{minipage}
    \caption{Posterior distribution of the decay temperature $\log_{10}(T_d/\GeV)$ and the wall tension $\log_{10}(\sigma_{\DW}/\GeV^3)$ with  $N_{\DW}=3$ and $6$ for hypothesis H6 in Table \ref{ABYS}.}
    \label{fig:ndwH6_td}
\end{figure*}

DWs with $N_{\DW}>2$ are also considered in our model, and we here list the search results for $N_{\DW}=3,4,5$ and $6$. First, we show the GW spectra corresponding to these $N_{\DW}$ values along with the free-spectra search of  the SGWB amplitude in Fig.~\ref{FS}. Then, for each $N_{\DW}$, we repeat the procedure of data analysis as done for the $N_{\DW}=2$ case. We summarize the detailed components of every hypothesis, the 1-$\sigma$ regions of parameters, and the Bayes factor between different hypotheses in Table~\ref{ABYS}. Figs. \ref{fig:ndwH2_td}-\ref{fig:ndwH6_td} show the posterior distributions of the parameters in different hypotheses described in the table, corresponding to the cases of the $N_{\DW}$ values of 3 and 6. As can be seen, DW networks with $N_{\DW}>2$ can also offer a good explanation for the strong CPL process, with the Bayes factors of H2 ranging from $10^{1.2}$ to $10^{1.8}$ against the null hypothesis H0. Moreover, results similar to the $N_{\DW}=2$ case also arise when we include the CPL process as a background component, no matter whether we maintain the auto-correlation between the pulsars or consider the CPL process as an astrophysical source from the SMBHB. The Bayes factors are all distributed around 1, meaning a disfavoured evidence for H3/4/6 against H1. It might be helpful to distinguish between DW networks with different $N_{\DW}$ values, with more observation data coming in the near future, but we can only give the constraints on the parameters for now. The 1-$\sigma$ C.L. limits of these parameters are all listed in Table \ref{ABYS}.

\begin{figure}[htp!]
    \centering
    \includegraphics[width=9cm]{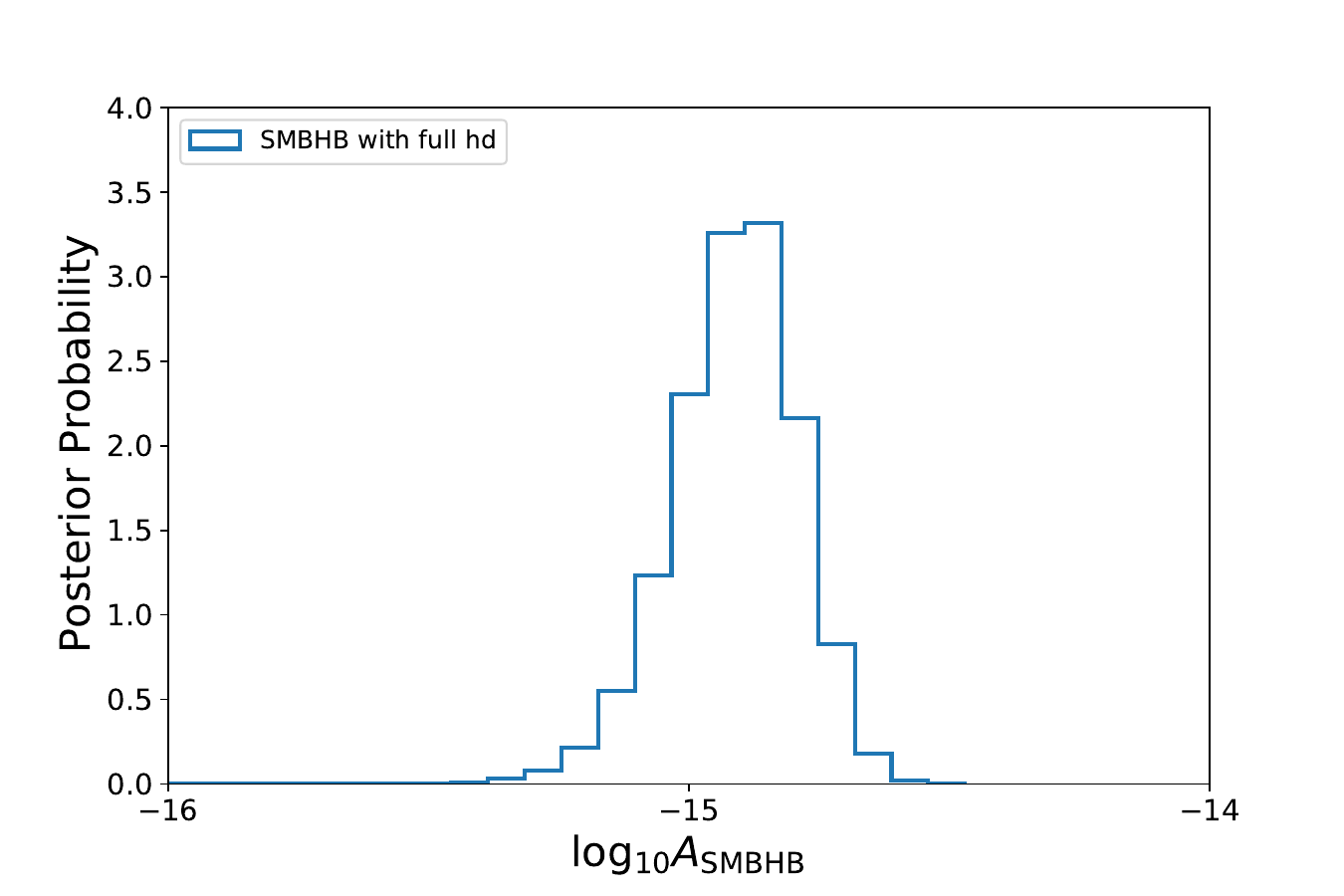}
    \caption{Posterior distribution of $A_{\mathrm{SMBHB}}$ of the H5 case in the Table\ref{ABYS}.}
    \label{fig:H5smbhb}
\end{figure}

To be complementary, we also reproduce the strong CPL signal using hypothesis H1. The CPL spectrum used in this paper is represented by
\begin{equation}
    S(f) = \frac{A_{\rm CPL}}{12\pi^2} \left( \frac{f}{yr^{-1}}\right)^{-\gamma_{\rm CPL}}yr^{3}.
\end{equation}
We obtain a 1-$\sigma$ region for the amplitude and spectra index: $\log_{10}A_{\rm CPL}=-14.48^{+0.62}_{-0.64}$ and $\gamma_{\rm CPL}=3.34^{+1.37}_{-1.53}$ with the Bayes factor of ${\rm BF}=10^{3.2}$. Then we made a trial to explain the CPL signal by using the solely SMBHB process (H5), and it turned out to be a good explanation with a large Bayes factor, ${\rm BF}=10^{3.3}$, against the null hypothesis H0. The posterior distribution for $\log_{10}A_{\rm SMBHB}$ is shown in Fig. \ref{fig:H5smbhb}. Then we also give the 1-$\sigma$ region of the amplitude: $\log_{10}A_{\rm SMBHB}=-14.89^{+0.10}_{-0.12}$. Note that the SGWB spectrum from the SMBHB used in this work is given by:
\begin{equation}
    S(f) = \frac{A_{\rm SMBHB}}{12\pi^2} \left( \frac{f}{yr^{-1}}\right)^{-\gamma_{\rm SMBHB}}yr^{3}.
\end{equation}
In this case, $\gamma_{\rm SMBHB}$ is fixed at $13/3$, and $A_{\rm SMBHB}$ is assumed to take the prior distribution of log-U $[-18, -14]$.

We notice that Ref.~\cite{Ferreira:2022zzo} made a similar search by using both the NANOGrav 12.5 years datasets \cite{Alam_2020} (NG12) and the International PTA Data Release 2 datasets \cite{Perera_2019} (IPTA DR2). Note that the PPTA DR2 we used in this work is independent of IPTA DR2. Besides, we also differ from them in the search strategies. 
The SGWB spectra used here are directly interpreted from the result of the large field simulation \cite{Hiramatsu2013}, and we extract out the corresponding spectra for every single value of $N_{\DW}$ instead of treating it as a continuous parameter. By doing this, we can remove the uncertainty of one additional parameter in searching. The searches in PPTA data yield similar results for $N_{\DW}=2$ to $6$ DW networks. This is as expected because they have similar GW spectra as can be seen in Table \ref{fitting}. To be concise, we have shown the $N_{\DW}=2$ case as a benchmark in the main text, while leaving the details of $N_{\DW}>2$ results in the appendix as shown above. 
We have made a more thorough search including a) a direct search for the SGWB in the data with only white and red noises included, which yields the best-fit parameter values of DW; b) include or not include the auto-correlation between the pulsar pars while taking the CPL as a systematic error, since the source of CPL still remains unknown; c) joint search for SGWB from the SMBHB and DW. As seen in \cite{Ferreira:2022zzo}, they only focused on parts a) and c). Besides, they set the HD overlap reduction function $\Gamma_{ab}=1$ to simplify the computation while we adopt the strict formula to be conservative

\section{The domain wall networks and $\Delta N_{\rm eff}$}

DW network can release its energy into two forms: GWs and free particles (\ie axions or scalar particles) that build the walls, with the quantitative relation $\rho_{\GW}/\rho_{\DW}=3/(8\pi)\cdot \epsilon \cdot \rho_{\DW}/\rho_c$. Taking the typical values $\sigma_{\DM}=10^{14}~\GeV^3$ and $T_d\sim 20$~MeV, we get $\rho_{\GW}/\rho_{c}\sim 10^{-5}$ and $\rho_{\DW}/\rho_{c}\sim 10^{-2}$ at $T_d$, which means that most of the energy is released into free particles. Furthermore, simulations show that those particles behave as cold matter with momentum comparable to the mass~\cite{Hiramatsu2013}. Consequently, for the parameter space we are interested in, such cold matter can dominate the Universe too early, which contradicts cosmological observations. Here, we consider one simple solution: those free particles can further decay into dark relativistic species (\ie dark radiation, DR) via their couplings with a dark sector~\cite{gonzalez2020ultralight,davoudiasl2020gravitational,berghaus2020thermal}.
In this picture, free particles decay and inject energy into the dark radiation. It causes an increase in the effective number of relativistic species, $\Delta N_{\rm eff}\equiv \rho_{\rm DR}/\rho_{\nu}^{\rm SM}$ where $\rho_{\nu}^{\rm SM}$ is the energy density of (a single flavor) standard model neutrinos. A model-independent calculation shows that our scenario gives
\beq\label{eq:N_eff}
&\Delta N_{\rm eff}
\simeq
\left(\frac{4}{11}\right)^{-\frac{4}{3}} \cdot \frac{4}{7} \cdot \frac{\rho_{\DW}(T_d)}{\rho_{c}(T_d)}  g_{*s}^{4/3}(T) g_{*s}^{-1/3}(T_d) \cdot F(\Gamma),  \\
& ~~~~~~~~~~~~~~~~ {\rm with~~} 
F(\Gamma) =
\int_{t_d}^t \frac{a(t')}{a(t_d)} {\rm e}^{-\Gamma (t'-t_d)}\Gamma dt'.
\eeq
$g_{*s}$ is the effective degree of freedom for entropy. $\Gamma$ is the decay rate from free particles to dark radiation. We require a large $\Gamma$ to complete the decay before the BBN epoch $\sim 1$~MeV to avoid the potential violation of BBN observations. If $\Gamma$ is sufficiently large ($\Gamma \gg t_d^{-1}$), $F(\Gamma)\simeq 1$ which returns to the case discussed in Ref.~\cite{Ferreira:2022zzo}. 
Furthermore, we expect that if the decay is not instant, in addition to alleviating the Hubble tension with the dark radiation, the residual free particles can also serve as dark matter. One can find various examples of such no-instant decay processes, for instance, in Ref.~\cite{Li:2020nah}). This issue is worthy of further study. Contrastingly, the network of long cosmic strings in the scaling regime can not alleviate the Hubble tension. The maximal contribution to $\Delta N_{\text{eff}}$ from its decay is less than $ 10^{-6}$, adhering to the limit established by LIGO's observations, where $\Omega_{\text{gw}}h^2 < 10^{-7}$~\cite{Dror:2019syi}. This underscores the significant role of the DW network in alleviating the Hubble tension.

Finally, we build a simple model as an example to allow free particles to further decay into dark radiation.
We consider a potential $V(\phi)= \lambda/4 (\phi^2 - \eta^2)^2$ which can form a $Z_2$ domain wall network. $\phi$ represents the scalar field that constitutes the wall, $\eta$ denotes the energy scale of spontaneous $Z_2$ symmetry breaking, and the domain wall tension, $\sigma_{\rm DW}$, is given by $\sigma_{\rm DW}=2\sqrt{2\lambda}\eta^3/3$. We suppose the $Z_2$ symmetry is explicitly broken by the term
\beq\label{eq:coupling_breaking}
V_{\rm breaking} = \kappa \eta \cdot \phi \chi^2.
\eeq
Here, $\chi$ signifies an ultralight scalar field that can be approximated as a classical field with the mean-field value $\left<\chi\right>$. This coupling term Eq.~\eqref{eq:coupling_breaking} fulfills a dual role. Firstly, it can induce wall decay through the bias potential $\Delta V = 2\kappa \eta^2 \left<\chi\right>^2$, which equates to the energy density $\sigma_{\rm DW}/t$ during the DW decay. Secondly, via this coupling term, the free particles resulting from wall decay will further rapidly decay into ultralight $\chi$ particles, which behave as dark radiation.

In our example model Eq.~\eqref{eq:coupling_breaking}, the interaction between dark radiation $\chi$ will be suppressed by the mediator mass, $m\sim \sqrt{\lambda \eta^2}$. However, this can be easily relaxed if we make $m$ light enough without violating the requirement that the domain wall network accounts for the SGWB.
The domain wall tension, $\sigma_{\rm DW} \sim \sqrt{\lambda}\eta^3
\sim m\eta^2$,
can still satisfy the preferred parameter space Eq.~(5) with appropriate values of $\lambda$ and $\eta$. Of course, the interaction between dark radiation does not need to be sourced by Eq.~\eqref{eq:coupling_breaking}. For example, one can introduce a self-coupling, e.g., $\lambda' \chi^4$,  to account for self-interaction between dark radiation, so Eq.~\eqref{eq:coupling_breaking} becomes
\beq
V =  \kappa \eta \cdot \phi \chi^2 + \lambda' \chi^4.
\eeq
Furthermore, Eq.~\eqref{eq:coupling_breaking} is just one interesting example that shows the possibility that the explicit soft symmetry breaking of $\phi$ and the self-interaction between dark radiation can be induced by a single term. However, Eq.~\eqref{eq:coupling_breaking} is not necessary, since one can introduce the two effects separately.

\begin{table*}[!htbp]
    \caption{Description of the hypotheses, Bayes factors, and parameters estimation of $N_{\DW}>2$.}
    \centering
    \setlength{\tabcolsep}{1.5mm}
    \renewcommand\arraystretch{1.8}
    \footnotesize
\begin{tabular}{|c|c|c|c|c|c|c|c|c|}
\hline
\multicolumn{2}{|c|}{\multirow{2}{*}{Hypothesis}} &\multirow{2}{*}{Pulsar Noise} & \multirow{2}{*}{CPL} & \multirow{2}{*}{SMBHB} & \multirow{2}{*}{DW Spectra} & \multicolumn{3}{c|}{Parameter Estimation ($68\%$ C.L.)}  \\ \cline{7-9}

\multicolumn{2}{|c|}{}  &  &  &  &    & \multicolumn{1}{c|}{$\log_{10}T_d /\GeV$, $\log_{10}\sigma_{\DW}/\GeV^3$} & 
$A_{\text{CPL}}, \gamma_{\text{CPL}},A_{\text{SMBHB}}$ ($T_d$) & Bayes Factors ($T_d$)
\\ \hline
\multicolumn{2}{|c|}{H0} & \checkmark &  &  &   &\multicolumn{1}{c|}{—} &\multicolumn{1}{c|}{—} & —\\ \hline
\multicolumn{2}{|c|}{H1}  & \checkmark & \checkmark &  & &  \multicolumn{1}{c|}{—} & $-14.48_{-0.64}^{+0.62}$, $3.34_{-1.53}^{+1.37}$, — &$10^{3.2}$ (/H0)
\\ \hline
\hline
\multicolumn{1}{|l|}{\multirow{4}{*}{H2}} & \multicolumn{1}{c|}{\multirow{1}{*}{$N_{\DW}=3$}} & \multicolumn{1}{c|}{\multirow{1}{*}{\checkmark}} & & &  \multicolumn{1}{c|}{\multirow{1}{*}{\checkmark(full HD)}} & \multicolumn{1}{c|}{$-1.34_{-0.56}^{+0.69}$, $14.31_{-1.35}^{+2.29}$} & \multicolumn{1}{c|}{\multirow{1}{*}{—}}  & \multicolumn{1}{c|}{$10^{1.8}$ (/H0)} \\ \cline{2-9}

\multicolumn{1}{|l|}{\multirow{4}{*}{}} & \multicolumn{1}{c|}{\multirow{1}{*}{$N_{\DW}=4$}} &  \multicolumn{1}{c|}{\multirow{1}{*}{\checkmark}} & & &  \multicolumn{1}{c|}{\multirow{1}{*}{\checkmark(full HD)}} & \multicolumn{1}{c|}{$-1.27_{-0.54}^{+0.67}$, $14.38_{-1.44}^{+2.31}$} &  \multicolumn{1}{c|}{\multirow{1}{*}{—}}  & \multicolumn{1}{c|}{$10^{1.5}$ (/H0)} \\ \cline{2-9}

\multicolumn{1}{|l|}{\multirow{4}{*}{}} & \multicolumn{1}{c|}{\multirow{1}{*}{$N_{\DW}=5$}} &  \multicolumn{1}{c|}{\multirow{1}{*}{\checkmark}} & & &  \multicolumn{1}{c|}{\multirow{1}{*}{\checkmark(full HD)}} & \multicolumn{1}{c|}{$-1.34_{-0.65}^{+0.70}$, $14.29_{-1.65}^{+2.28}$} &  \multicolumn{1}{c|}{\multirow{1}{*}{—}}  & \multicolumn{1}{c|}{$10^{1.2}$ (/H0)} \\ \cline{2-9}

\multicolumn{1}{|l|}{\multirow{4}{*}{}} & \multicolumn{1}{c|}{\multirow{1}{*}{$N_{\DW}=6$}} &  \multicolumn{1}{c|}{\multirow{1}{*}{\checkmark}} & & &  \multicolumn{1}{c|}{\multirow{1}{*}{\checkmark(full HD)}} & \multicolumn{1}{c|}{$-1.32_{-0.69}^{+0.69}$, $14.30_{-1.67}^{+2.29}$} &  \multicolumn{1}{c|}{\multirow{1}{*}{—}}  & \multicolumn{1}{c|}{$10^{1.2}$ (/H0)} \\

\hline
\hline
\multicolumn{1}{|l|}{\multirow{4}{*}{H3}} & \multicolumn{1}{c|}{\multirow{1}{*}{$N_{\DW}=3$}} & \multicolumn{1}{c|}{\multirow{1}{*}{\checkmark}} &\multicolumn{1}{c|}{\multirow{1}{*}{\checkmark}} & &  \multicolumn{1}{c|}{\multirow{1}{*}{\checkmark(full HD)}} & \multicolumn{1}{c|}{$>-1.31,<16.50$} & \multicolumn{1}{c|}{$-14.63_{-0.97}^{+0.67},3.34_{-1.67}^{+1.49}$, —} 
& \multicolumn{1}{c|}{$0.98$ (/H1)}\\ \cline{2-9}

\multicolumn{1}{|l|}{\multirow{4}{*}{}} &  \multicolumn{1}{c|}{\multirow{1}{*}{$N_{\DW}=4$}} & \multicolumn{1}{c|}{\multirow{1}{*}{\checkmark}} &\multicolumn{1}{c|}{\multirow{1}{*}{\checkmark}} & &  \multicolumn{1}{c|}{\multirow{1}{*}{\checkmark(full HD)}}  & \multicolumn{1}{c|}{$>-1.22,<16.51$} & \multicolumn{1}{c|}{$-14.67_{-0.99}^{+0.66},3.41_{-1.65}^{+1.47}$, —}  & \multicolumn{1}{c|}{$1.00$ (/H1)} \\ \cline{2-9}

\multicolumn{1}{|l|}{\multirow{4}{*}{}} &  \multicolumn{1}{c|}{\multirow{1}{*}{$N_{\DW}=5$}} & \multicolumn{1}{c|}{\multirow{1}{*}{\checkmark}} &\multicolumn{1}{c|}{\multirow{1}{*}{\checkmark}} & &  \multicolumn{1}{c|}{\multirow{1}{*}{\checkmark(full HD)}}  & \multicolumn{1}{c|}{$>-1.28,<16.47$} & \multicolumn{1}{c|}{$-14.63_{-0.89}^{+0.65},3.38_{-1.62}^{+1.43}$, —}  &  \multicolumn{1}{c|}{$0.97$ (/H1)}\\ \cline{2-9}

\multicolumn{1}{|l|}{\multirow{4}{*}{}} &  \multicolumn{1}{c|}{\multirow{1}{*}{$N_{\DW}=6$}} & \multicolumn{1}{c|}{\multirow{1}{*}{\checkmark}} &\multicolumn{1}{c|}{\multirow{1}{*}{\checkmark}} & &  \multicolumn{1}{c|}{\multirow{1}{*}{\checkmark(full HD)}}  & \multicolumn{1}{c|}{$>-1.30,<16.53$} & \multicolumn{1}{c|}{$-14.58_{-1.09}^{+0.68},3.18_{-1.65}^{+1.57}$, —}  & \multicolumn{1}{c|}{$0.97$ (/H1)} \\
\hline
\hline

\multicolumn{1}{|l|}{\multirow{4}{*}{H4}} & \multicolumn{1}{c|}{\multirow{1}{*}{$N_{\DW}=3$}} & \multicolumn{1}{c|}{\multirow{1}{*}{\checkmark}} &\multicolumn{1}{c|}{\multirow{1}{*}{\checkmark}} & &  \multicolumn{1}{c|}{\multirow{1}{*}{\checkmark(no-auto HD)}} & \multicolumn{1}{c|}{$>-0.95,<16.52$} & \multicolumn{1}{c|}{$-11.11_{-0.13}^{+0.08}$, $0.34_{-0.21}^{+0.37}$, —}  & \multicolumn{1}{c|}{$0.89$ (/H1)} \\ \cline{2-9}

\multicolumn{1}{|l|}{\multirow{4}{*}{}} &  \multicolumn{1}{c|}{\multirow{1}{*}{$N_{\DW}=4$}} & \multicolumn{1}{c|}{\multirow{1}{*}{\checkmark}} &\multicolumn{1}{c|}{\multirow{1}{*}{\checkmark}} & &  \multicolumn{1}{c|}{\multirow{1}{*}{\checkmark(no-auto HD)}}& \multicolumn{1}{c|}{$>-0.97,<16.46$} & \multicolumn{1}{c|}{$-11.07_{-0.09}^{+0.05}$, $0.65_{-0.18}^{+0.27}$, —}  & \multicolumn{1}{c|}{$0.89$ (/H1)} \\ \cline{2-9}
 
\multicolumn{1}{|l|}{\multirow{4}{*}{}} &  \multicolumn{1}{c|}{\multirow{1}{*}{$N_{\DW}=5$}} & \multicolumn{1}{c|}{\multirow{1}{*}{\checkmark}} &\multicolumn{1}{c|}{\multirow{1}{*}{\checkmark}} & &  \multicolumn{1}{c|}{\multirow{1}{*}{\checkmark(no-auto HD)}} & \multicolumn{1}{c|}{$>-0.23,<16.31$} & \multicolumn{1}{c|}{$-13.12_{-0.13}^{+0.13}$, $3.62_{-0.28}^{+0.32}$, —}  & \multicolumn{1}{c|}{$0.93$ (/H1)} \\ \cline{2-9}
 
\multicolumn{1}{|l|}{\multirow{4}{*}{}} &  \multicolumn{1}{c|}{\multirow{1}{*}{$N_{\DW}=6$}} & \multicolumn{1}{c|}{\multirow{1}{*}{\checkmark}} &\multicolumn{1}{c|}{\multirow{1}{*}{\checkmark}} & &  \multicolumn{1}{c|}{\multirow{1}{*}{\checkmark(no-auto HD)}} & \multicolumn{1}{c|}{$>-1.06,<16.40$} & \multicolumn{1}{c|}{$-13.08_{-0.16}^{+0.17}$, $3.87_{-0.38}^{+0.38}$, —}  & \multicolumn{1}{c|}{$0.89$ (/H1)} \\ 
\hline
\hline
\multicolumn{2}{|c|}{H5} & \checkmark &  & \checkmark &  &  \multicolumn{1}{c|}{ —} & —, —, $-14.89_{-0.12}^{+0.10}$ &$10^{3.3}$ (/H0)
\\ \hline
\hline

\multicolumn{1}{|l|}{\multirow{4}{*}{H6}} & \multicolumn{1}{c|}{\multirow{1}{*}{$N_{\DW}=3$}} & \multicolumn{1}{c|}{\multirow{1}{*}{\checkmark}} & &\multicolumn{1}{c|}{\multirow{1}{*}{\checkmark}} &  \multicolumn{1}{c|}{\multirow{1}{*}{\checkmark(full HD)}} & \multicolumn{1}{c|}{$>-1.00,<16.47$} & \multicolumn{1}{c|}{—, —, $-14.92_{-0.16}^{+0.11}$}  &  \multicolumn{1}{c|}{$0.91$ (/H5)}\\ \cline{2-9}

\multicolumn{1}{|l|}{\multirow{4}{*}{}} & \multicolumn{1}{c|}{\multirow{1}{*}{$N_{\DW}=4$}} & \multicolumn{1}{c|}{\multirow{1}{*}{\checkmark}} & &\multicolumn{1}{c|}{\multirow{1}{*}{\checkmark}} &  \multicolumn{1}{c|}{\multirow{1}{*}{\checkmark(full HD)}} & \multicolumn{1}{c|}{$>-1.07,<16.46$} & \multicolumn{1}{c|}{—, —, $-14.92_{-0.16}^{+0.11}$}  & \multicolumn{1}{c|}{$0.90$ (/H5)} \\ \cline{2-9}

\multicolumn{1}{|l|}{\multirow{4}{*}{}} &  \multicolumn{1}{c|}{\multirow{1}{*}{$N_{\DW}=5$}} & \multicolumn{1}{c|}{\multirow{1}{*}{\checkmark}} & &\multicolumn{1}{c|}{\multirow{1}{*}{\checkmark}} &  \multicolumn{1}{c|}{\multirow{1}{*}{\checkmark(full HD)}}& \multicolumn{1}{c|}{$>-1.05,<16.44$} & \multicolumn{1}{c|}{—, —, $-14.93_{-0.16}^{+0.12}$} &\multicolumn{1}{c|}{$0.91$ (/H5)} \\ \cline{2-9}

\multicolumn{1}{|l|}{\multirow{4}{*}{}} & \multicolumn{1}{c|}{\multirow{1}{*}{$N_{\DW}=6$}} & \multicolumn{1}{c|}{\multirow{1}{*}{\checkmark}} & &\multicolumn{1}{c|}{\multirow{1}{*}{\checkmark}} &  \multicolumn{1}{c|}{\multirow{1}{*}{\checkmark(full HD)}}& \multicolumn{1}{c|}{$>-1.11,<16.46$} & \multicolumn{1}{c|}{—, —, $-14.93_{-0.18}^{+0.12}$} & \multicolumn{1}{c|}{$0.92$ (/H5)} \\ 
\hline
\end{tabular}\label{ABYS}
\end{table*}

\bibliography{arxiv_v3}

\bibliographystyle{apsrev}

\end{document}